\documentclass[aps,prb,showpacs,twocolumn,floatfix]{revtex4}
\usepackage{graphicx}
\usepackage[usenames]{color}
\usepackage{bbm}
\usepackage{array}
\begin{document}

\title{Diffusion and desorption of $\mathrm{SiH_3}$ on hydrogenated $\mathrm{H:Si\left(100\right)-(2\times 1)}$
from first principles.}
\author{Michele Ceriotti}
\email{michele.ceriotti@phys.chem.ethz.ch}
\altaffiliation{present address: Computational Science, Department of Chemistry and Applied Biosciences, ETH Z{\"u}rich,
USI Campus, Via Giuseppe Buffi 13, CH-6900 Lugano, Switzerland}

\author{Marco Bernasconi}
\affiliation{Dipartimento di Scienza dei Materiali and CNISM, 
 Universit\`a di  Milano-Bicocca, Via R. Cozzi 53, I-20125, Milano, Italy}

\begin{abstract}

We have studied diffusion pathways  of a silyl radical adsorbed on
the hydrogenated  $\mathrm{Si\left(100\right)}$-($2\times 1)$ 
surface within density-functional theory.
The process is of interest for the  growth of crystalline silicon by
plasma-enhanced chemical vapor deposition (PECVD).
Preliminary searches for migration mechanisms have been performed using
metadynamics simulations. Local minima and transition states  
have been further refined by using the Nudged-Elastic-Band method.
Barriers for diffusion from plausible adsorption sites as low as 0.35 eV 
have been found, but  trap states have also been spotted, leading to a
more stable configuration, with escape barriers of 0.80 eV.
Diffusion through weakly bound  physisorbed states is also possible with
very low activation barriers ($<$50 meV).
However, desorption mechanisms 
(either as $\mathrm{SiH_3}$ 
or as $\mathrm{SiH_4}$) from physisorbed or more strongly bound adsorption
configurations turn out to have activation energies similar to diffusion
barriers. Kinetic Monte Carlo simulations based on ab-initio activation energies
show that the silyl radical diffuses at most by a few lattice spacing before desorbing
at temperatures in the range 300-1000 K. 
\end{abstract}
\maketitle

\section{Introduction}

Plasma-enhanced chemical vapor deposition (PECVD) from silane is a widespread                
technique employed to grow thin films of amorphous silicon. \cite{donahue1985jap}
High growth rates are made possible by the deposition of reactive radicals
produced in the plasma, as opposed to conventional CVD where the less
reactive silane is directly adsorbed at the growing surface.
By suitably controlling the energy
of the ions while keeping a high density of reactants, a related technique,
called low-energy plasma-enhanced chemical vapor deposition (LEPECVD),
\cite{rosenblad1998tsf,rosenblad1998tsfb,rosenblad1998jcg,kummer2002}
has introduced the possibility 
 to obtain device-quality epitaxial
films of crystalline silicon or silicon-germanium
alloys, at temperatures much lower (600 $^o$C) 
than 
those necessary for conventional, thermal chemical vapor deposition (800 $^o$C).
$\mathrm{SiH_3}$ is supposed to be the most abundant radical species in the
plasma discharge in most of the experimental conditions\cite{abelson1993apa,vonkeudell1999prb}.
A description of the interactions of the silyl radical with the crystalline
$\mathrm{Si\left(100\right)}$ surface is then of
great interest to model the epitaxial growth 
at conditions of both PECVD and LEPECVD. 

The silyl radical can interact with the surface giving rise to different
processes: it can remove a hydrogen from a saturated dimer via an Eley-Rideal
mechanism, or it might adsorb on the surface, where it  would further evolve by diffusing,
decomposing with hydrogen release at the surface 
or by desorbing (either as $\mathrm{SiH_3}$ or  $\mathrm{SiH_4}$). 

The fate of the silyl radical depends on the degree of surface hydrogenation at the landing site.
At low hydrogen coverage the silyl can adsorb on a silicon dangling bond and 
 decompose easily (into SiH$_2$ for instance \cite{kang}).
The sticking probability is supposed to decrease at higher hydrogen  coverage.
The decomposition of the adsorbed radical is also supposed to be strongly hindered by the lack of
free silicon surface atoms to which H might be transferred.
Diffusion of the adsorbed silyl from hydrogen rich to hydrogen poor regions might be a
conceivable route for SiH$_3$ decomposition and insertion into the growing film.
Fast diffusion of SiH$_3$ adsorbed on the hydrogenated H:Si(100)-(2x1) surface has been
recently predicted by first principles calculations \cite{bakos2006,valipa2005prl}.
An equally fast diffusion of silyl radicals at the hydrogenated surface of amorphous silicon is
also often invoked as the reason behind the high smoothness of the amorphous films
grown by PECVD \cite{valipa2005prl}.
Although 
several processes involving the adsorbed silyl radical at the Si(100) surface 
  have already been investigated in literature, previous theoretical
studies  always considered \emph{a-priory} guess of reaction 
pathways, a procedure which might overlook unexpected mechanisms for diffusion and reaction
of the adsorbed 
species.\cite{gupta2002,bakos2006,bakos2005ieee,ramalingam1998,walch2000,bakos2005jcp,bakos2005cpl,valipa2005prl} 

In this paper,  we  investigate further the diffusion of SiH$_3$ adsorbed on the H:Si(100)-(2x1) surface
by making use of the 
ab-initio  metadynamics technique\cite{laio2002,iannuzzi2003},
a new simulation tool which allows for extensive search of diffusion pathways.
Local minima visited during metadynamics trajectories have been then optimized 
 and  migration barriers between different minima
 further refined by  the Nudged Elastic Band (NEB) method\cite{henkelman2000a,henkelman2000b}.
Competitive mechanisms for diffusion and desorption have been investigated by
Kinetic Monte Carlo simulations which provide the average diffusion length the silyl radical
travels before eventually desorbing.

After a brief description  of our theoretical framework in Section \ref{sec:compu-details}, 
we present in Section \ref{sec:meccanismi} our results on the  diffusion mechanisms for 
the adsorbed  silyl radical, on  
the existence of trap states which hinder the radical 
mobility and on desorption processes which compete with diffusion.
In Section III  we report the results of the Kinetic Monte Carlo simulations
based on the ab-initio activation energies which has allowed  estimating the effect
of possible errors in the calculated activated energies on the fate of the silyl radical.
Sec. IV is devoted to our conclusions.

\section{\label{sec:compu-details}Computational details}
Calculations have been performed within the framework of Density Functional 
Theory (DFT) with a gradient corrected  exchange and correlation energy 
functional (PBE) \cite{perdew1996}  as implemented in the codes PWSCF\cite{pwscf}
for geometry optimizations and NEB\cite{henkelman2000a} calculations 
and CPMD\cite{cpmd} for Car-Parrinello\cite{carparrinello1985} metadynamics  
simulations \cite{laio2002,iannuzzi2003}.
Norm-conserving \cite{troullier1991} and ultrasoft \cite{vanderbilt1990}
pseudopotentials have  been used 
for silicon and
hydrogen, respectively.
Kohn-Sham orbitals are expanded  in  plane waves up to  kinetic energy cutoff of 25 Ry.
Selected calculations have been repeated with a norm conserving pseudopotential for H as well
and a kinetic cutoff of 30 Ry.

\begin{figure}
\caption{\label{fig:2x1}(color online) The slab model of the $\mathrm{H:Si\left(100\right)-(2\times 1)}$ surface,
from a top a) and side b) views.}
\centering\includegraphics[width=0.9\columnwidth]{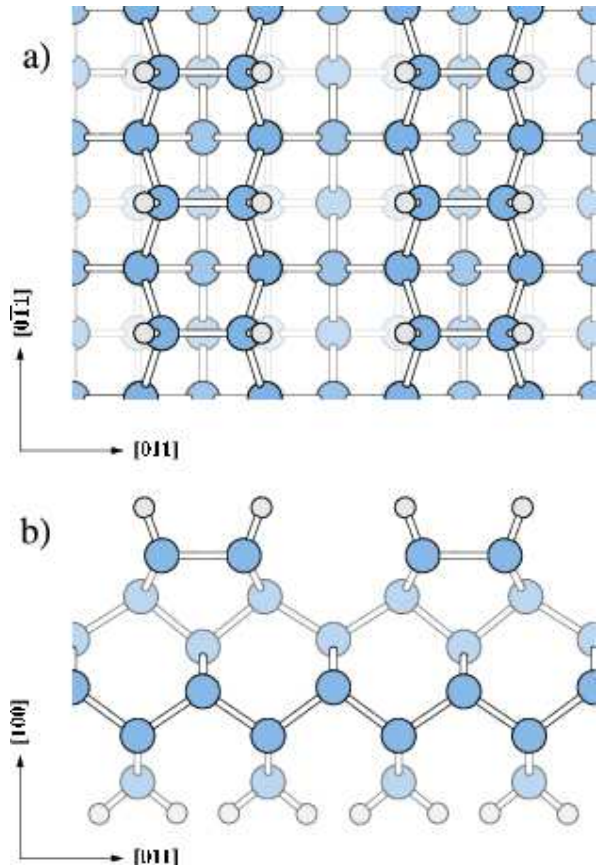}
\end{figure}

As  preliminary tests of our framework, we have calculated the reaction enthalpies for
some decomposition reactions of disilane (Si$_2$H$_6$)
 within a cubic supercell of edge
 15.3 \AA. 
Both PBE and BLYP \cite{becke1988,lee1988} have been used for these benchmark 
calculations on  molecular systems.
Geometries have been optimized until the maximum residual force
on every atom was less than 0.01 eV/\AA. 
 Formation enthalpies have been calculated taking into
account the vibrational contribution to the quantum partition function
in the harmonic regime, with normal modes
obtained, in turn, by diagonalization of the dynamical
matrix, built from the numerical derivatives of the forces with respect to
finite atomic displacements (0.015~\AA).
Rotational and translational contributions to reaction enthalpies have been added
in the classical limit.
Results for PBE  and
BLYP functionals are compared in 
Table~\ref{tab:molecole} with experimental data and previous quantum-chemical
calculations. The results are in good agreement with experiments and
higher levels of theory,  PBE performing overall better than  BLYP functional. 
For  reactions involving homolithic bond breaking such as the decomposition
$\rm{Si_2H_6\rightarrow 2SiH_3}$, 
spin  unrestricted calculations allowing spin polarization are mandatory to reproduce  the
correct reaction enthalpies. 
Neglect of spin polarization (spin restricted calculations) introduces errors as large as 1 eV 
(cf. Table I) in the reaction energies. 
All the surface calculations have been then performed in a spin unrestricted framework (LSD-PBE).

\begin{table}
\caption{\label{tab:molecole} 
Reaction energies ($\Delta U_0$) and enthalpies ($\Delta H$, in eV) calculated for some gas-phase
decomposition reactions of disilane. Results  obtained with PBE and BLYP 
functionals are compared to experimental data and previous quantum-chemical results.}
\begin{ruledtabular}
\begin{tabular}[c]{l c c c}
$\mathrm{Si_2H_6\rightarrow}$&$\mathrm{SiH_2 + SiH_4}$&$\mathrm{H_2 + Si_2H_4}$&$\mathrm{SiH_3+SiH_3}$\\\hline
$\Delta U_0$ PBE	& 2.559	& 2.169	& 3.147	\\
$\Delta U_0$ PBE\footnote{Spin restricted calculation}
& 	& 	& 4.158	\\
$\Delta U_0$ BLYP	& 2.213	& 2.099	& 2.984	\\
$\Delta U_0$ MP2\footnote{MP2/6-311++g** results from Ref.\cite{hu2003}}
	& 2.436	& 2.375	& 3.182	\\\hline
$\Delta H$ PBE
		& 2.445	& 2.033	& 3.028	\\
$\Delta H$ BLYP\footnote{Vibrational contributions to enthalpy from PBE normal modes.}
		& 2.098	& 1.963	& 2.865	\\
$\Delta H$ MP2\footnotemark[2]
		& 2.355	& 2.064	& 3.048	\\
$\Delta H$ B3LYP\footnote{from Ref.\cite{tonokura2002}}
	& 2.259	& 2.025	& 3.265	\\
$\Delta H$ Exp.\footnotemark[4]
		& 2.355	& 2.021	& 3.326	\\
\end{tabular}
\end{ruledtabular}
\end{table}

\begin{figure}
\caption{\label{fig:nebtest} Minimum energy path obtained by NEB optimization for the migration from the minimum (a) to the 
minimum (b) of Figure~\ref{fig:minimi}.
The energy profile with respect to a
suitably defined reaction coordinate $q$ is reported.
 Path A is  optimized with the $\Gamma$-point only.
Path B corresponds to the total energies           
obtained with one special $k$-point in the surface BZ for the geometries of path A.
Path C has been fully optimized by
 using one special $k$-point.}
\centering\includegraphics[width=0.9\columnwidth]{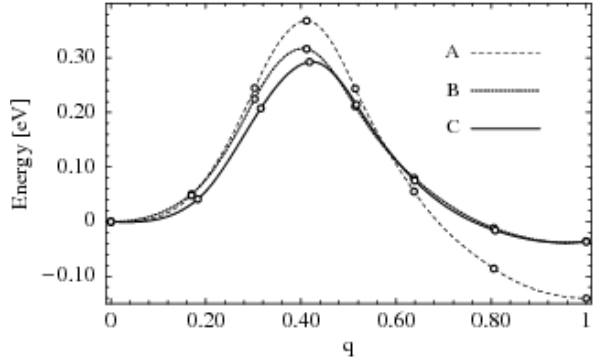}
\end{figure}

The H:Si(001)(2x1) surface is modeled in a slab geometry with 3D periodic boundary conditions
(see Fig.\ref{fig:2x1}).
The theoretical equilibrium lattice parameter obtained from a bulk calculation with the
k-point sampling corresponding to the supercell $\Gamma$-point is 0.7 $\%$ larger than the
experimental one, while the theoretical value at full convergence in
BZ integration is 0.6 $\%$ shorter than the experimental one.
Thus, in the slab calculation we chose the experimental lattice parameter 
(5.43  {\AA} \cite{hartwigm1994}) as an average value of the theoretically lattice constants that can 
be obtained with different supercell  sizes and k-point sampling.
In a previous work \cite{nostroprb}, we checked that
by changing  the lattice parameter by
 0.7 $\%$, the adsorption energy of the silyl radical
on the clean Si(001)-(2x1) surface
 changes by less than 10 meV.
The periodically repeated slabs are separated by vacuum, 11~\AA~ wide.
The  slab has six silicon layers, each containing 12 atoms.
The hydrogenated top surface, reconstructed in the (2x1) geometry, contains
six hydrogenated silicon dimers.
The bottom surface of the slab  is saturated by symmetric $\mathrm{SiH_2}$
groups. 
The SiH$_2$ groups and the underlying silicon layer were kept fixed 
at the ideal bulk positions. 
Only the supercell $\Gamma$-point has been  considered in Brillouin Zone
sampling for geometry optimizations.
We checked the convergence of our results with respect to Brillouin Zone integration
by optimizing selected geometries with the  special $k$-point
 $\left(\frac{1}{4}, \frac{1}{4}\right)$ (in crystallographic coordinates
\cite{cunningham1974}). Calculations with large surface supercell (three rows of five dimers each)
have also been performed for selected systems.

To uncover possible diffusion pathways of the silyl radical, we have
made use of the metadynamics technique 
which allows large barriers to be overcome in an affordable simulation time (few picoseconds)
within ab-initio molecular dynamics simulations.\cite{laio2002,iannuzzi2003} 
The method is based on a
coarse-grained, non-Markovian dynamics in the manifold spanned by few reaction coordinates 
biased by a history-dependent potential which drives the system towards the lowest saddle point.
 The main assumption is that the reaction path could be described on the manifold
of few collective variables (CV)  $S_{\alpha}(\{{\bf R}_I\})$, function of the ionic coordinates ${\bf R}_I$.
The Lagrangian $\mathcal{L}_0$ of the system is then modified, 
introducing an history-dependent biasing potential, which affects the dynamics so as 
to discourage the system
from remaining in the region  already visited and pushes it over the lowest
energy barrier towards a new equilibrium basin.
Many variations over these basic principles have been explored; in this paper we  use the
 straightforward approach of Ref.\cite{laio2005}, which simply introduces a repulsive potential 
($\mathcal{L}=\mathcal{L}_0-g\left(\{\mathbf{R}_I\},t\right)$) built from the 
superposition of Gaussians centered at points previously visited by the trajectory in CV space,
which acts directly on the ionic  coordinates $\mathbf{R}_I$ as

\begin{equation}
g\left(\{\mathbf{R}_I\},t\right)=w \sum_{t_j<t}
	\exp\left[-\frac{
			\sum_{\alpha=1}^n\left(S_{\alpha}\left(\{\mathbf{R}_I\left(t\right)\}\right)-
					S_{\alpha}\left(\{\mathbf{R}_I\left(t_j\right)\}\right)\right)^2
			}{ 2\Delta s^2 }
		\right]
\label{eq:metapotential}
\end{equation}

To study the diffusion of the adsorbed silyl radical, we have chosen as collective variables the
$(x,y)$ surface position of the silyl silicon atom (two CVs). 
The Gaussian hills parameters (Eq.~\ref{eq:metapotential}) have been 
chosen as $\Delta s=0.32$ a.u. and $w=0.11$ eV; a new hill is added each time the 
trajectory reaches a point $2\Delta s$ far from the previous Gaussian in CV space, or at 
worst every 250 time steps.
We have performed  
Car-Parrinello metadynamics simulations 
 with a time-step of 6 a.u., effective mass for  electrons of 600 a.u.,
and deuterium mass for hydrogen. 
Constant temperature (300~K) on ions is enforced by a Nos\`e-Hoover thermostat.\cite{hoover1985} 

Although  metadynamics allows computing activation free-energies from a finite temperature
simulation, long simulation time (with small Gaussian height $w$) are needed to obtain
accurate estimates of activation free energies. Actually, our metadynamics simulations are aimed
at just obtaining a good starting guess for the transformation path.
Then, the geometry and activation energy of the transition state identified along
the dynamical trajectory have  been
further refined by using a standard   technique with fixed end points,
the Nudged Elastic Band (NEB) method.\cite{henkelman2000a} 
Climbing image and
variable springs \cite{henkelman2000b} have been used thoroughly (but for 
processes with a barrier below 0.10 eV, where a constant spring constant of 0.6 a.u. has 
been used), with spring constants 
$k_{max}=0.6$ a.u. and $k_{min}=0.3$ a.u.. A  minimization scheme has been applied until 
the residual total forces acting on each image in direction perpendicular to the path 
was less than 0.05 eV. 
We have relaxed the Minimum Energy Path 
(MEP) using $\Gamma$-only calculations,  then we have performed a self-consistent 
electronic structure optimization along the MEP using one special $k$-point. 
Full optimization of the MEP geometry with one special point 
(cf. Fig.~\ref{fig:nebtest} for a selected process) introduces changes in the activation energies 
of the order of a few tens of meV  with respect to the calculation with one special 
point on the gamma-point geometry.

To obtain some figures describing 
qualitatively the behavior of the radical, and to test how much the fate of the
radical is
 affected by variations of the calculated activation energies within the 
 uncertainties of DFT (up to 0.1~eV). 
 we have performed Kinetic Monte Carlo (KMC)
\cite{lebowitz} simulation based on the reaction scheme  from metadynamics and
NEB calculations.

KMC is a simulation technique which permits to attain large length and time scales, 
relying on the knowledge of the rates for all the relevant reaction mechanisms
between local minima. 
The dynamics is performed stepwise, choosing at every step one
of the possible mechanism with a probability proportional to its rate, and 
incrementing the simulation time by the appropriate time step
following the scheme of Ref. \cite{lebowitz}. 

\section{Results}

\subsection{\label{sec:meccanismi}Mechanisms for diffusion and desorption}
The  starting geometry of our metadynamics simulations corresponds
to a SiH$_3$ radical adsorbed in the configuration of Fig.~\ref{fig:minimi}(a)
previously identified 
as a possible adsorption geometry 
from first principles static calculations \cite{valipa2005prl}
and from direct simulations of the impact of the silyl on the H:Si(100)2x1 surface \cite{nostroprb}.
In fact, we have shown by targeted ab-initio MD simulations that the SiH$_3$ sticks on the
surface in the configuration of Fig.~\ref{fig:minimi}(a) once it impinges on this site
with a translational kinetic energy in the range 0.1-0.2 eV. \cite{nostroprb}

\begin{figure}
\caption{\label{fig:meta-sih3}(color online) Four snapshots from the metadynamics
trajectory of silyl diffusion on H:Si(100)-(2x1): a) $t=0$ ps, b) $t=2.1$ ps, c) $t=3.5$ ps, d) $t=10$ ps. 
}
\large\bfseries
\begin{tabular}[c]{@{}>{\raggedleft}m{0.05\columnwidth}@{}>{\centering}m{0.95\columnwidth}@{}}
&\tabularnewline
a) & \includegraphics[width=0.75\columnwidth]{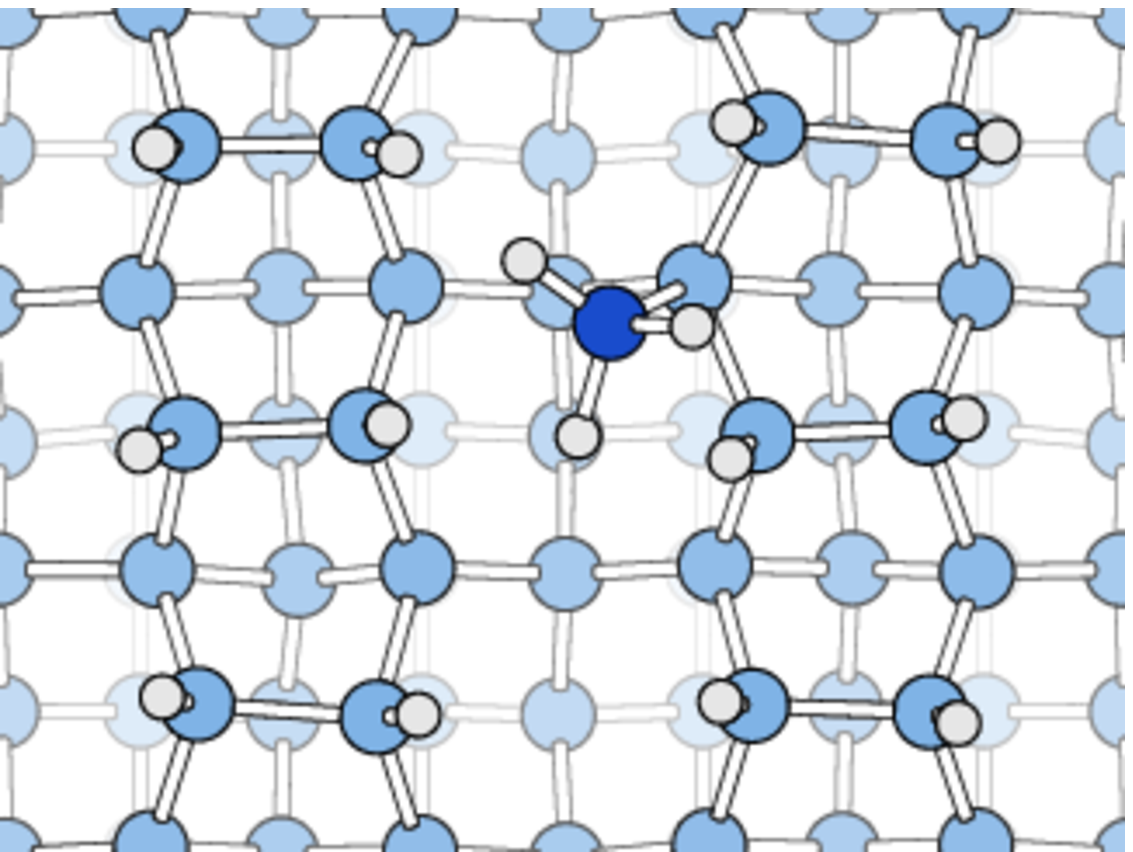}\tabularnewline
&\tabularnewline
b) & \includegraphics[width=0.75\columnwidth]{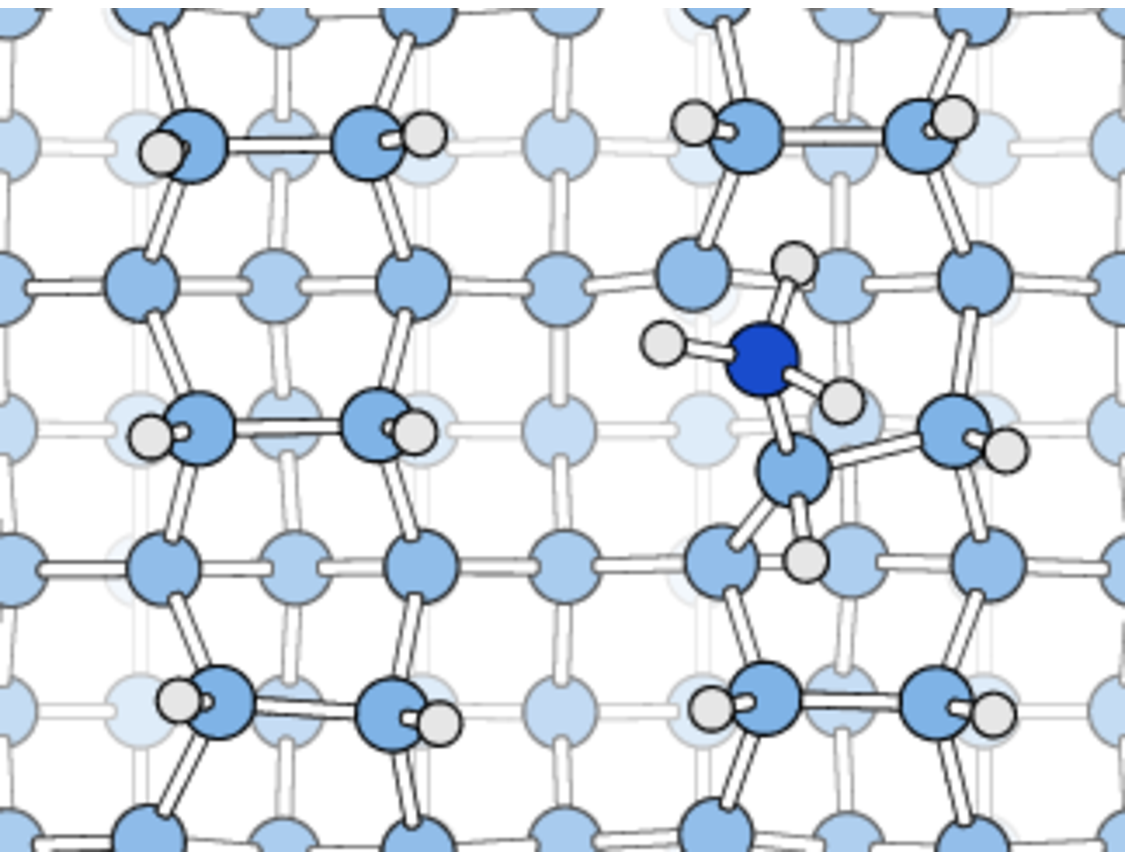}\tabularnewline
&\tabularnewline
c) & \includegraphics[width=0.75\columnwidth]{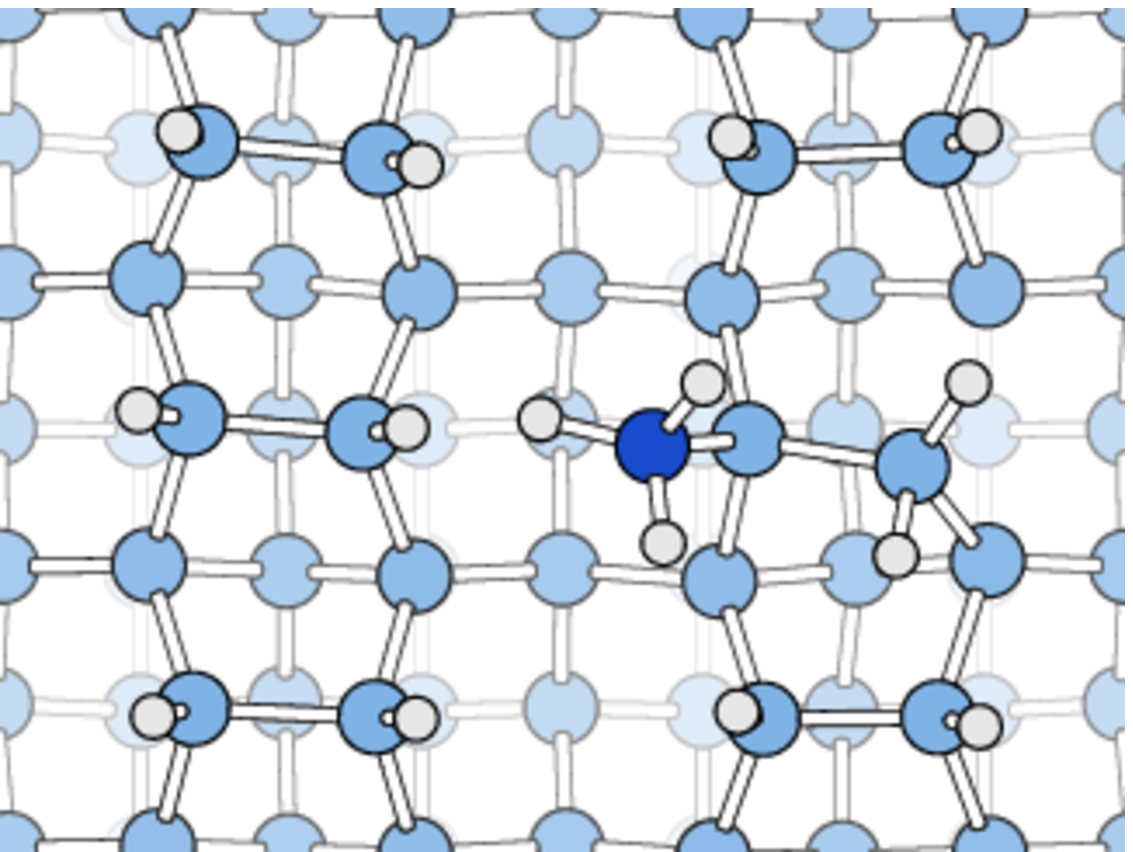}\tabularnewline
&\tabularnewline
d) & \includegraphics[width=0.75\columnwidth]{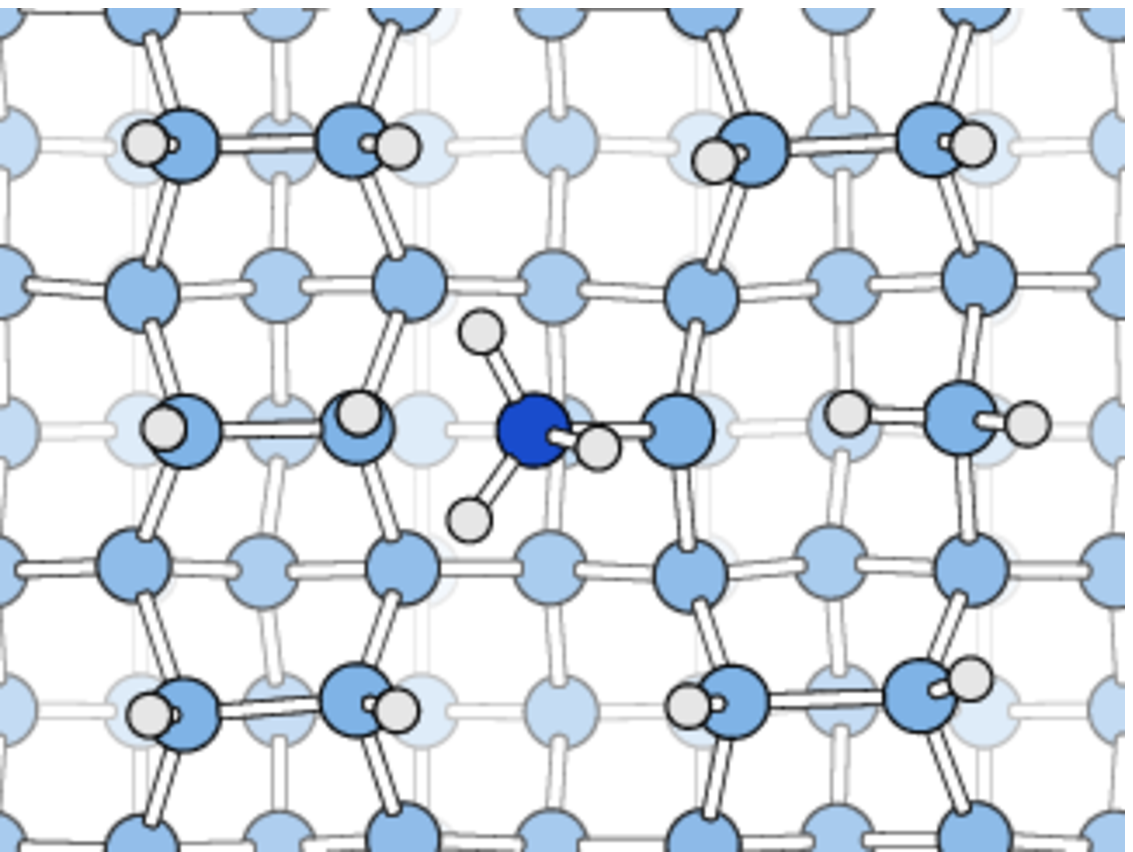}\tabularnewline
\end{tabular}
\end{figure}

In a metadynamics run 10 ps long, we have observed the diffusion of the SiH$_3$ radical via configurations not
previously considered in literature such as those 
 sketched in Figure~\ref{fig:meta-sih3}.
Indeed, inspection on the molecular dynamics trajectories has allowed identifying
several possible local minima and migration paths for the silyl radical which we have then
refined by geometry optimization and NEB calculations as described below.

\begin{table}
\caption{\label{tab:testcelle} 
Energies (eV) of  the  local minima shown in Figure~\ref{fig:minimi}.
The zero of energy corresponds to  SiH$_3$ far from the surface, as
computed from free surface and isolated silyl within the same supercell used
to model the adsorbed silyl.
$\mathrm{SiH_4} + db$ indicates the reaction energy of silane formation from a
silyl radical abstracting a surface hydrogen atom.
Column {\bf A} refers to the results that are
used throughout the paper, while other columns refer to various tests we 
performed on selected structures by changing $k$-points sampling and supercell size.
Namely, column {\bf A} refers to calculations with 
6 dimers supercell and  energies computed with one special $k$-point
on geometries optimized with the  $\Gamma$ point only. Column {\bf B} refers to the same calculations as in 
{\bf A} with energies as well as geometries computed with the  $\Gamma$ point only.
 Column {\bf C} refers to calculations with 6 dimers supercell with energies and geometries optimized with
 one special $k$-point.  Column {\bf D} refers to calculations with a 15 dimers supercell,  energies and geometries optimized
with  $\Gamma$ point only.   Column {\bf E} refers to  calculations as in column {\bf B} but without
spin polarization, i.e. within a spin-restricted framework (LDA-PBE instead of LSD-PBE).}
\begin{ruledtabular}
\begin{tabular}[c]{l c c l c c}
	& {\bf A}
 	& {\bf B}
 	& {\bf C}
	& {\bf D}
 	& {\bf E} \\ \hline
a)			& -0.15		& -0.23		& -0.14		& -0.16		& -0.64		\\
b)			& -0.18		& -0.38		&      		& -0.23		&  		\\
c)			& -0.15		& -0.32		&      		& -0.20		&  		\\
d)			& -0.25		& -0.45		& -0.26		& -0.30		&  		\\
e)			& -0.34		& -0.43		&      		& -0.36		&  		\\
f)			& -0.60		& -0.67		&      		& -0.61		&  		\\
g)			& -0.37		& -0.54		&      		&     		& -0.64		\\
h)			& -0.06 	& -0.07		&      		&      		& 		\\
i)			& -0.04   	& -0.05		&      		&      		& 		\\
j)			& -0.07 	& -0.08		&      		&      		& 		\\
$\mathrm{SiH_3}$	&  0.00		&  0.00		&  0.00		&  0.00		&  0.00		\\
$\mathrm{SiH_4+db}$	& -0.50 	& -0.52		& -0.50 	&      		& -0.80		\\

\end{tabular}
\end{ruledtabular}
\end{table}

\begin{figure}
\caption{\label{fig:minimi}(color online) Sketches of ten different local minima for  
$\mathrm{SiH_3}$ adsorbed on  $\mathrm{H:Si\left(100\right)-(2\times 1)}$ surface.
Side views of (a) and (i) structures are also reported at the bottom to illustrate  the difference
in geometry
between strongly bound and physisorbed minima.
The structures (e) and (g) are not perfectly symmetric with respect to the 
dimer plane, so that two corresponding mirror structures exist; this small asymmetry is due
to the balance between the interaction of the $\mathrm{SiH_3}$ hydrogen atoms with the surface 
hydrides, and the barrier for interconversion between the mirror configurations is likely
to be very low. For what concerns diffusion, the two mirror images can be regarded
as a single, symmetric structure.
}
\large\bfseries
\begin{tabular}[c]{@{}>{\raggedleft}m{0.05\columnwidth}@{}>{\centering}m{0.45\columnwidth}@{}>{\raggedleft}m{0.05\columnwidth}@{}>{\centering}m{0.45\columnwidth}@{}}
&&&\tabularnewline
(a) & \includegraphics[width=0.4\columnwidth]{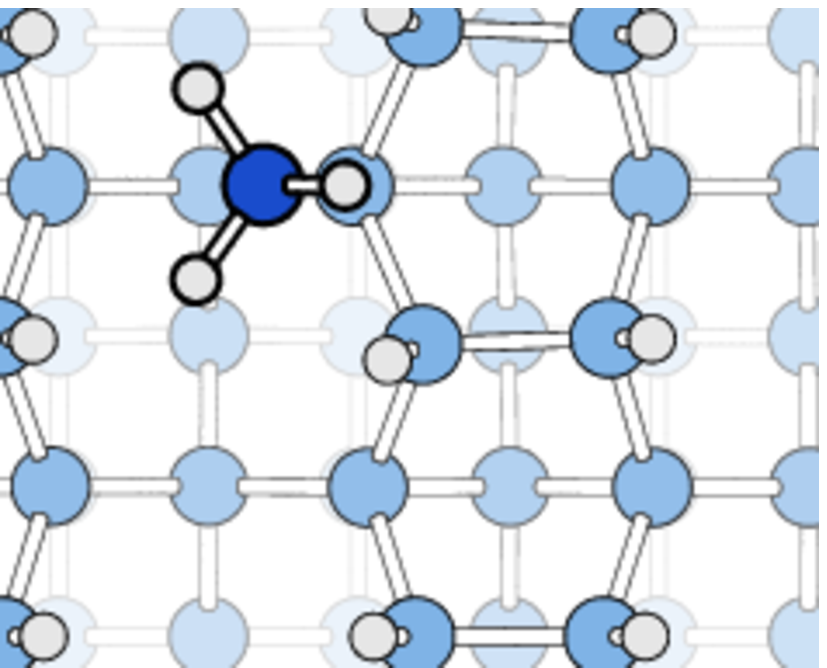} &
(b) & \includegraphics[width=0.4\columnwidth]{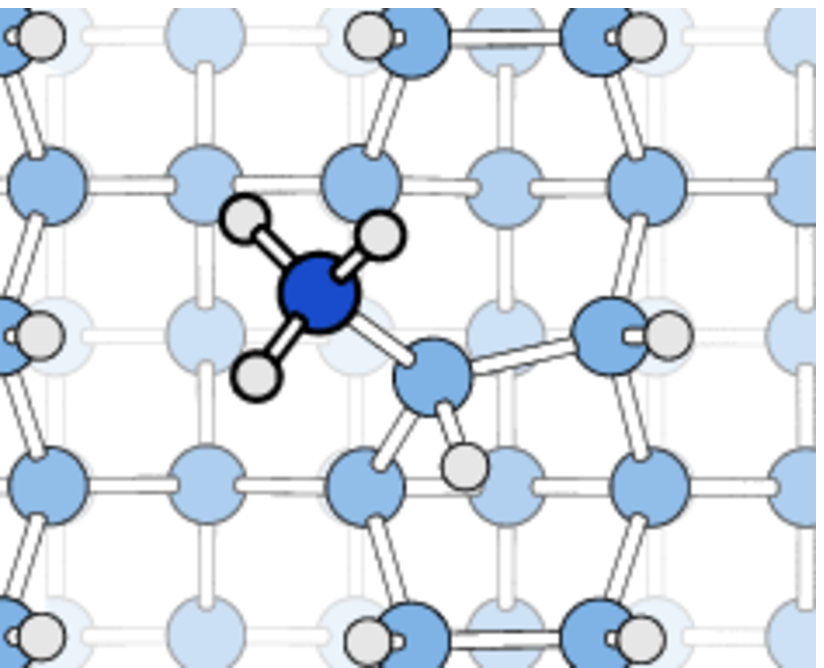} \tabularnewline
&&&\tabularnewline
(c) & \includegraphics[width=0.4\columnwidth]{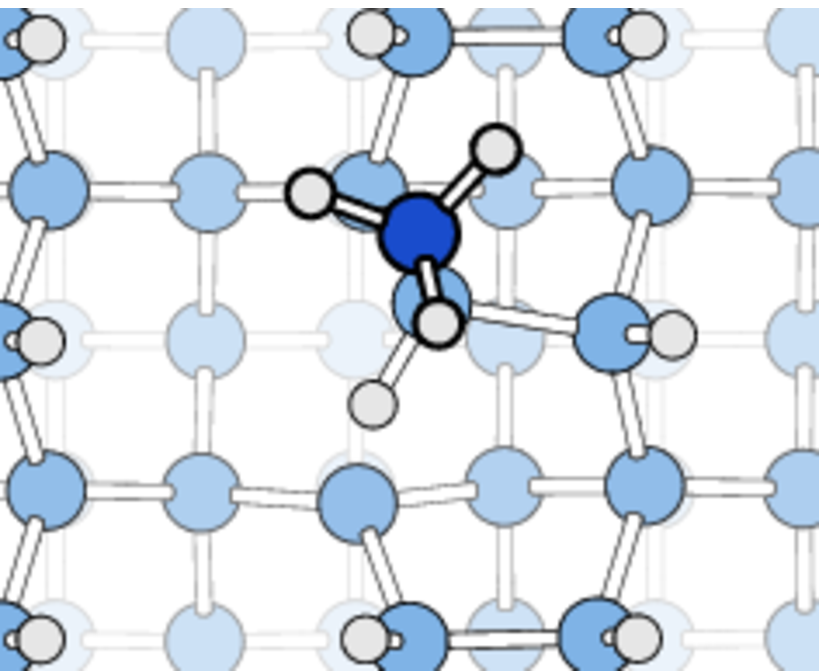} & 
(d) & \includegraphics[width=0.4\columnwidth]{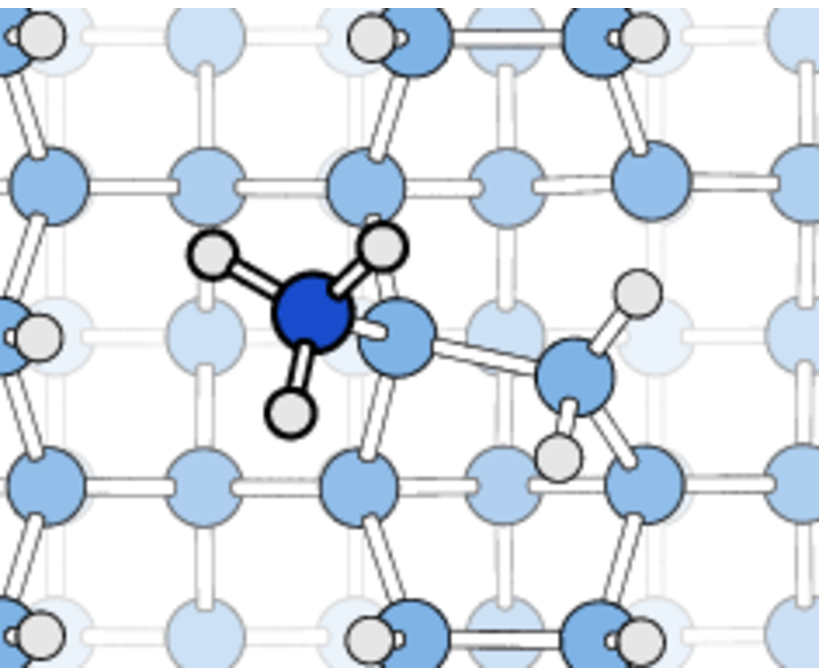} \tabularnewline
&&&\tabularnewline
(e) & \includegraphics[width=0.4\columnwidth]{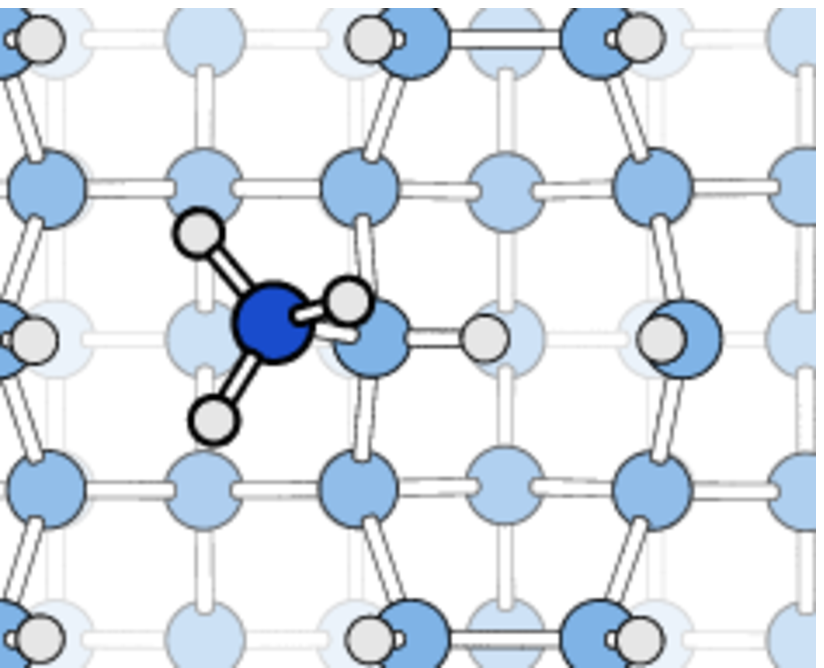} & 
(f) & \includegraphics[width=0.4\columnwidth]{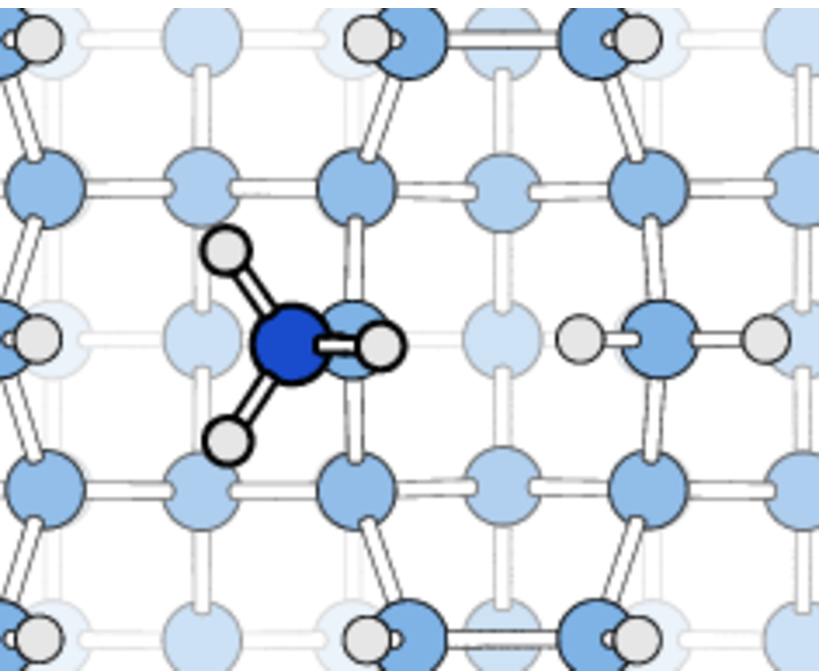} \tabularnewline
&&&\tabularnewline
(g) & \includegraphics[width=0.4\columnwidth]{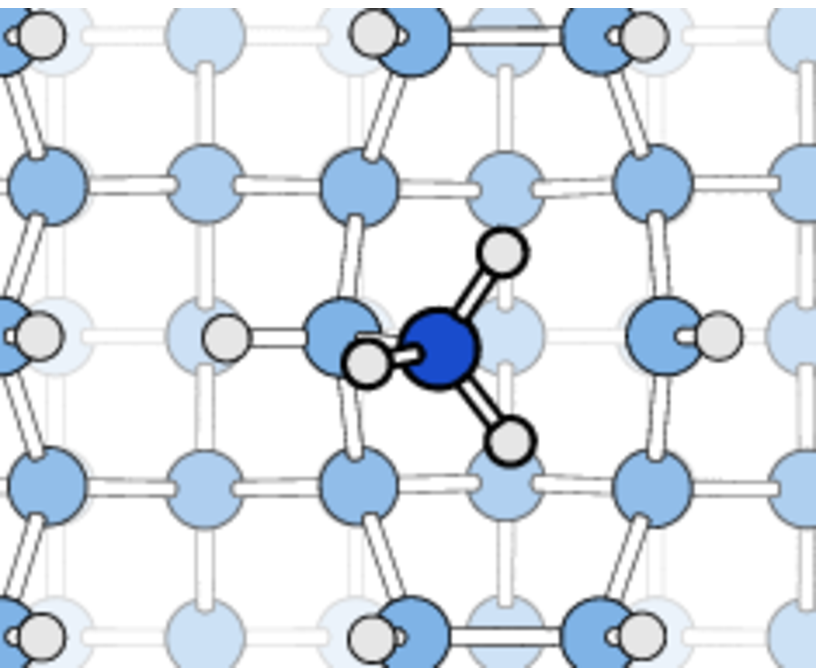} & 
(h) & \includegraphics[width=0.4\columnwidth]{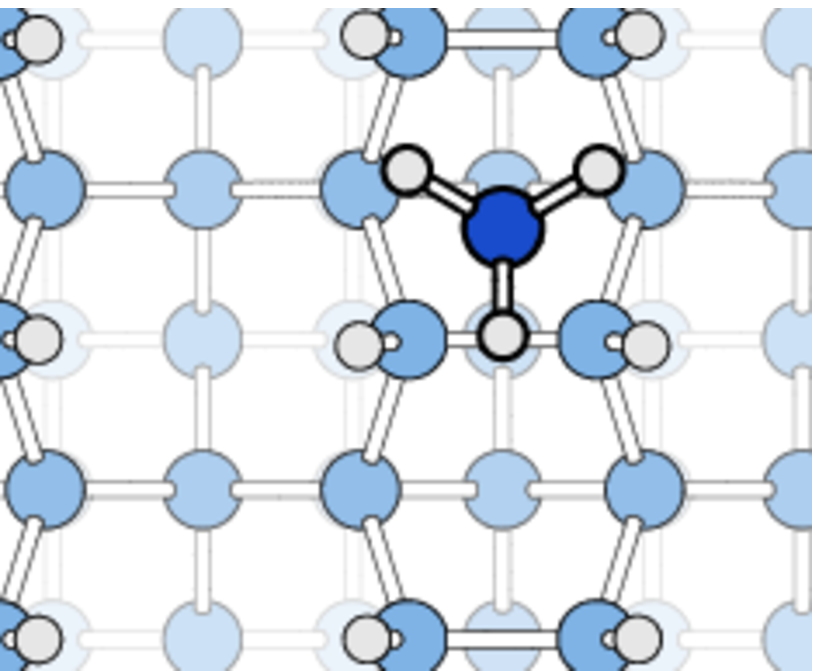} \tabularnewline
&&&\tabularnewline
(i) & \includegraphics[width=0.4\columnwidth]{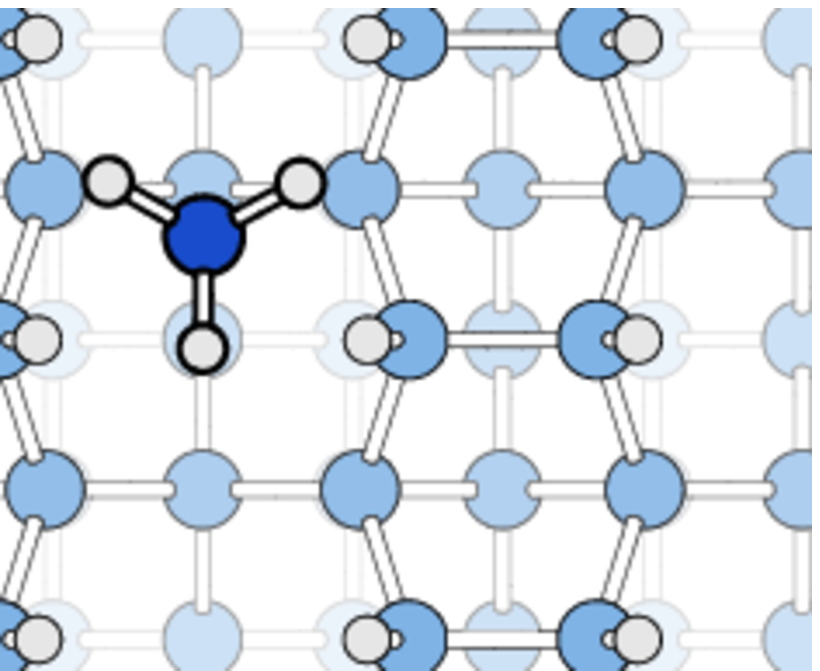} & 
(j) & \includegraphics[width=0.4\columnwidth]{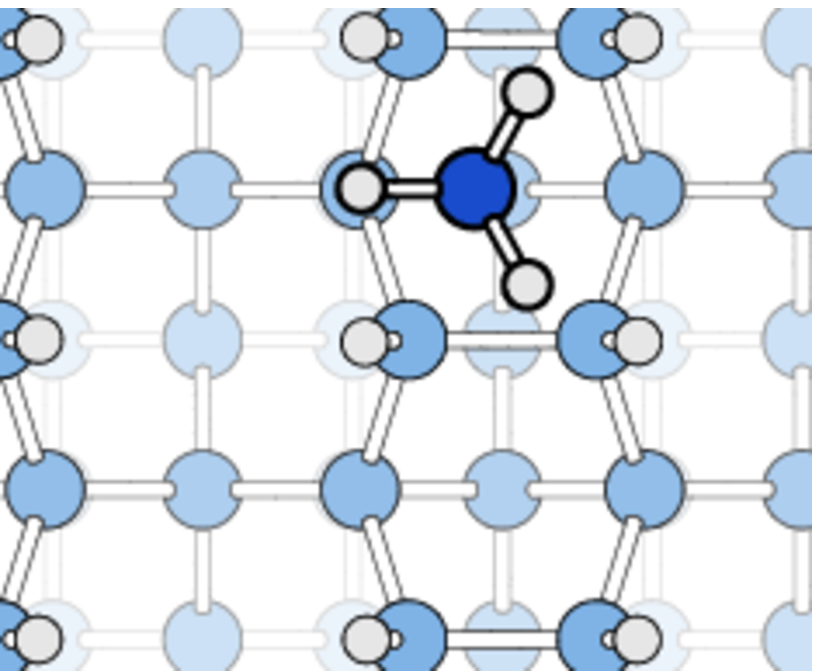} \tabularnewline
(a) & \includegraphics[width=0.4\columnwidth]{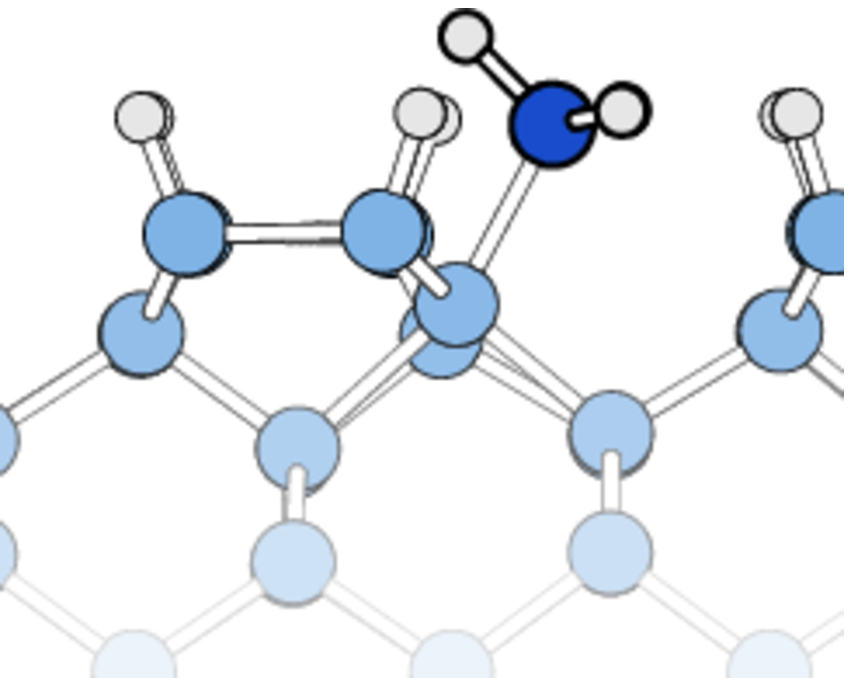} & 
(i) & \includegraphics[width=0.4\columnwidth]{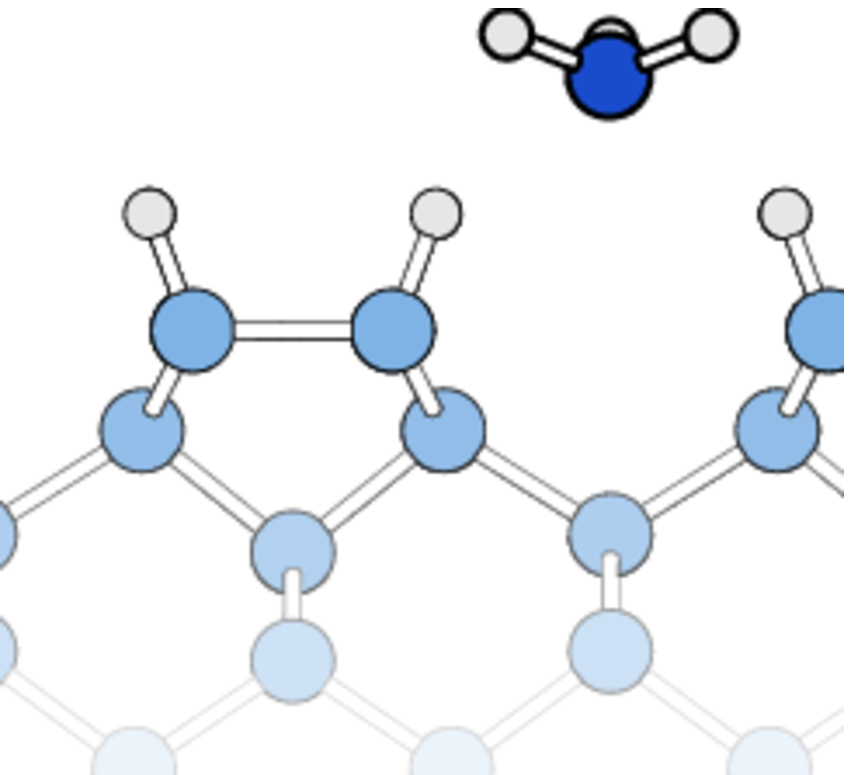} \tabularnewline
\end{tabular}
\end{figure}

Sketches of the local minima resulting from geometry optimization of 
the metadynamics snapshots are shown in Fig.
~\ref{fig:minimi} (in the
following, we will refer to ``configuration (x)'' as ``the configuration
sketched in Figure~\ref{fig:minimi} (x)'') along with pictures  of other structures 
taken from literature, or  guessed by analogy with  those identified in 
the metadynamics simulation. The energy of such minima are reported
in Table~\ref{tab:testcelle}.
As anticipated in section \ref{sec:compu-details} we have checked the convergence of our
results with respect to the size of the supercell and to the  k-points sampling 
of the Brillouin Zone (BZ).
Tests with more than one special  $k$-point have revealed  minute changes in energy differences
(the desorption energy from the (a) site changes from 0.147 eV to 0.154 eV  by using
one or four special $k$-points in the total energy calculation on geometries optimized with
$\Gamma$-point only).
On the other hand, to achieve reasonable accuracy in  activation energies, it is sufficient to use 
    one special point  on geometries optimized with gamma-point only (cf. Figure~\ref{fig:nebtest}). 

The calculated activation energies for the various migration processes among the different local minima 
are listed in 
Table~\ref{tab:attivaz}, and a schematic (and crowded) summary of the mechanisms examined is
drawn in Figure~\ref{fig:casino}.

Large disagreement is found for the adsorption energy 
 with respect to some ab-initio results 
previously reported in literature.
For instance in  Ref.\cite{bakos2006} an adsorption energy of 0.63 and 0.75 eV are reported 
for silyl in configurations (a) and (g), while  we obtain the values 0.15 and 0.35 eV, respectively. 
Similarly, the energy gain for the removal of a hydrogen by a 
silyl radical is as high as 0.86 eV in Ref.\cite{bakos2005ieee} to be compared with our result of
0.35 eV. 
A possible source of discrepancy  might be
the use of a spin restricted framework in Ref. \cite{bakos2005ieee,bakos2006}.
Indeed, by repeating our calculation with no spin polarization we have obtained adsorption and
reaction energies similar to those reported in previous works 
(column {\bf E} in  Table~\ref{tab:testcelle}).
As discussed in Sec \ref{sec:compu-details},  spin unrestricted framework is, however, recommended to properly describe an
open shell system like the silyl radical, especially in the gas phase.

\begin{figure*}
\caption{\label{fig:spinloc}(color online)  
Spin polarization 
($m\left(\mathbf{r}\right)=\rho_{\uparrow}\left(\mathbf{r}\right)-\rho_{\downarrow}\left(\mathbf{r}\right)$)
for the four configurations (a), (b), (f) and (j)  of Figure~\ref{fig:minimi}, 
is shown in the four corresponding panels.
Isosurfaces at $m=\pm 10^{-3}$ a.u. are reported, with 
 colors depending on the spin sign.
In the three structures, $\int{\mathrm{d}^3\mathbf{r}\left|m\left(\mathbf{r}\right)\right|}$
is approximately 1.2 au, corresponding to the unpaired electron introduced by the silyl, 
plus some polarization effects.
Structure (a) shows the most pronounced spin delocalization, with the unpaired electron
shared between the silyl silicon and two surface silicon atoms.}
\large\bfseries
\begin{tabular}[c]{@{}>{\raggedleft}b{0.05\textwidth}@{}>{\centering}b{0.45\textwidth}@{}>{\raggedleft}b{0.05\textwidth}@{}>{\centering}b{0.45\textwidth}@{}}
(a)\\~\\~\\~\\~\\~\\~\\~\\~\\~ \vfill & \includegraphics[width=0.4\textwidth]{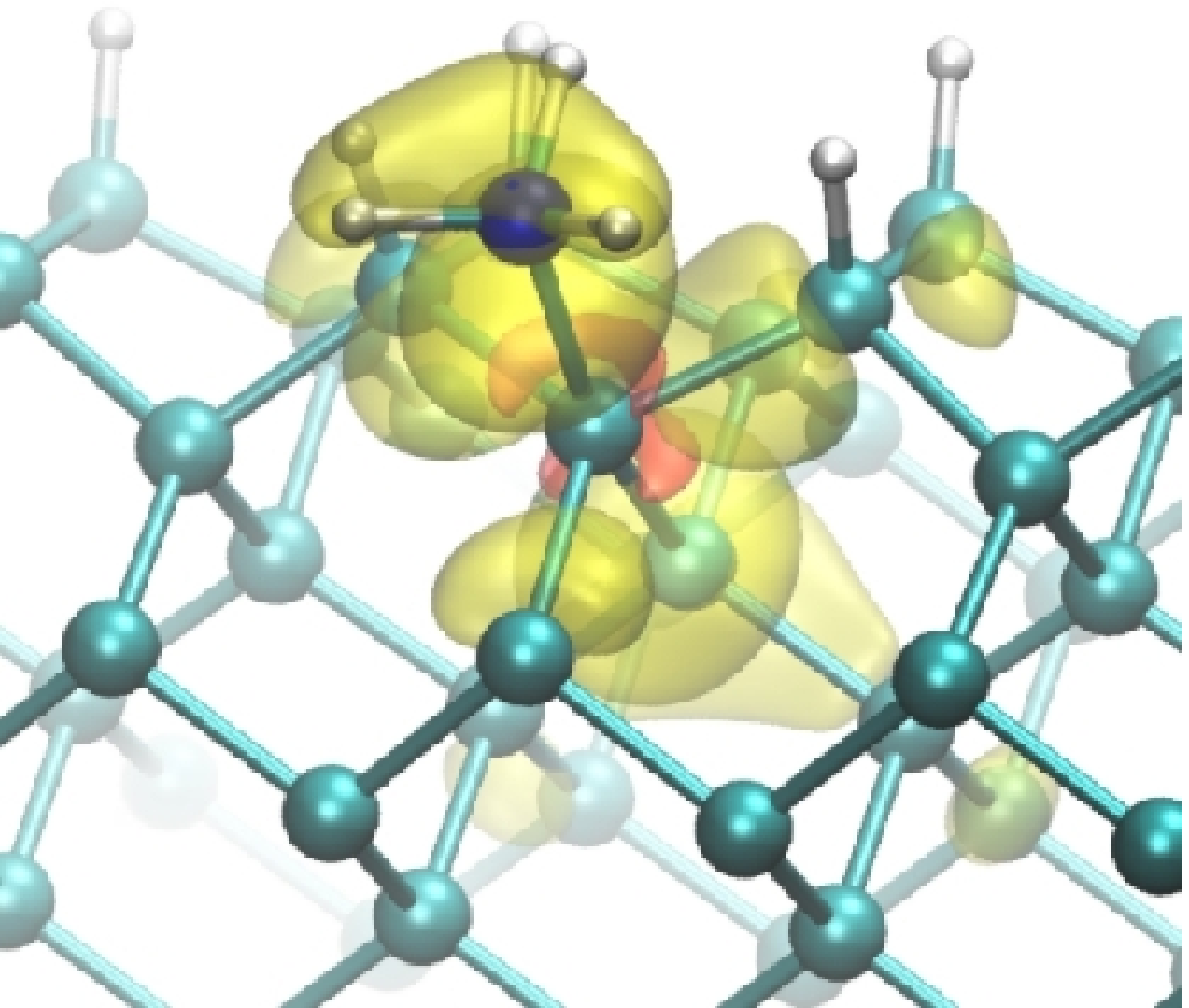}&
(b)\\~\\~\\~\\~\\~\\~\\~\\~\\~ \vfill & \includegraphics[width=0.4\textwidth]{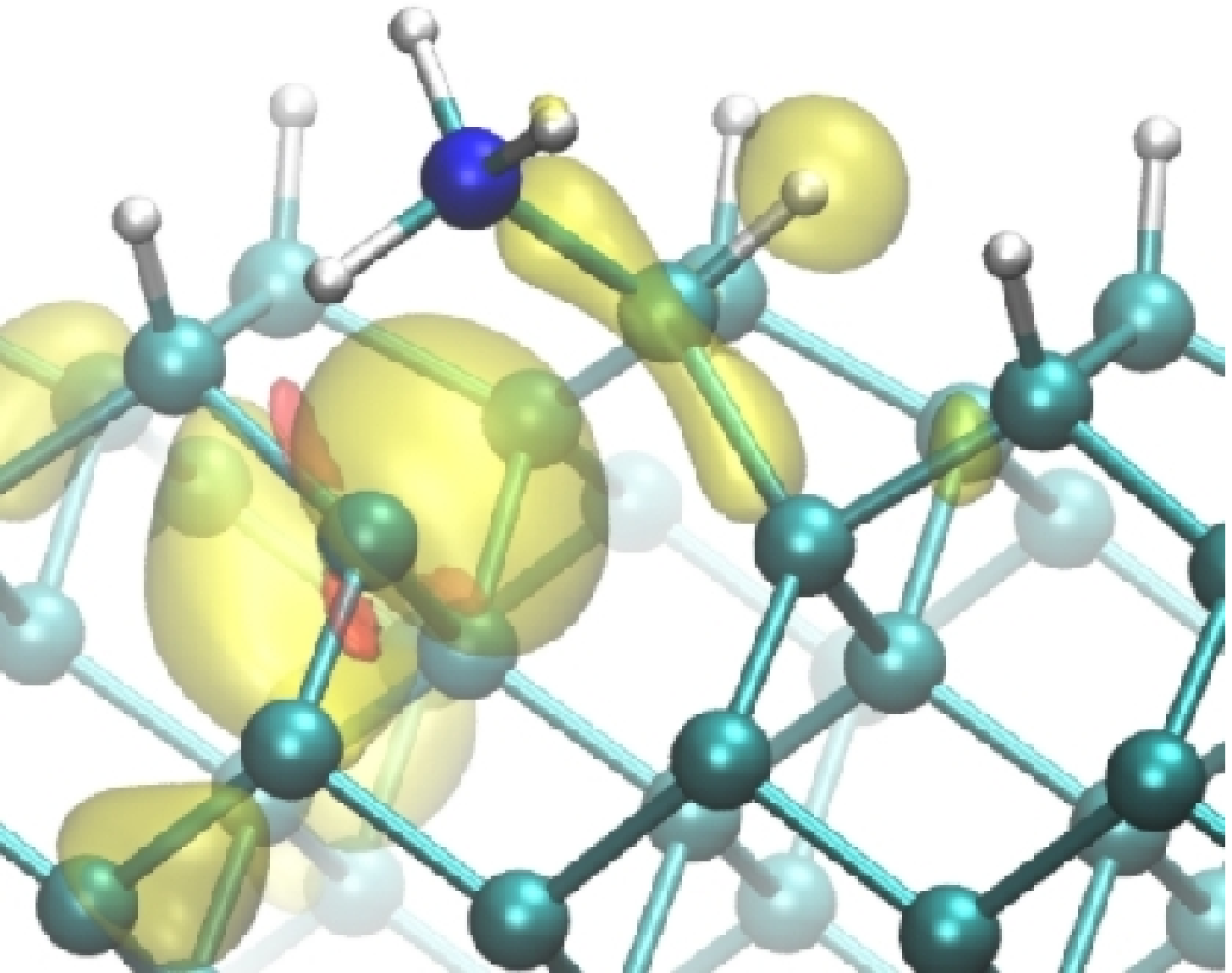}\tabularnewline
(f)\\~\\~\\~\\~\\~\\~\\~\\~\\~ \vfill & \includegraphics[width=0.4\textwidth]{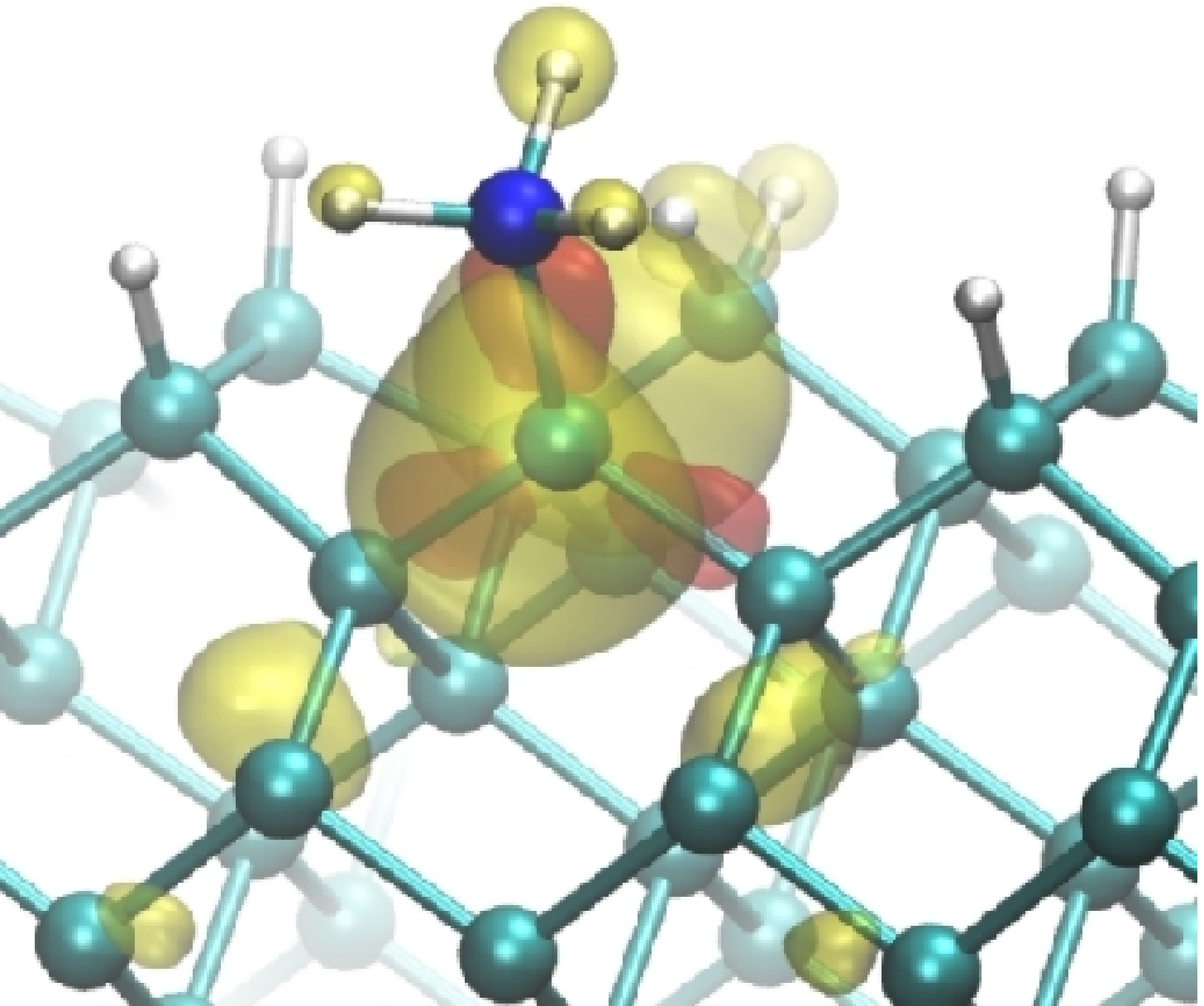}&
(j)\\~\\~\\~\\~\\~\\~\\~\\~\\~ \vfill & \includegraphics[width=0.4\textwidth]{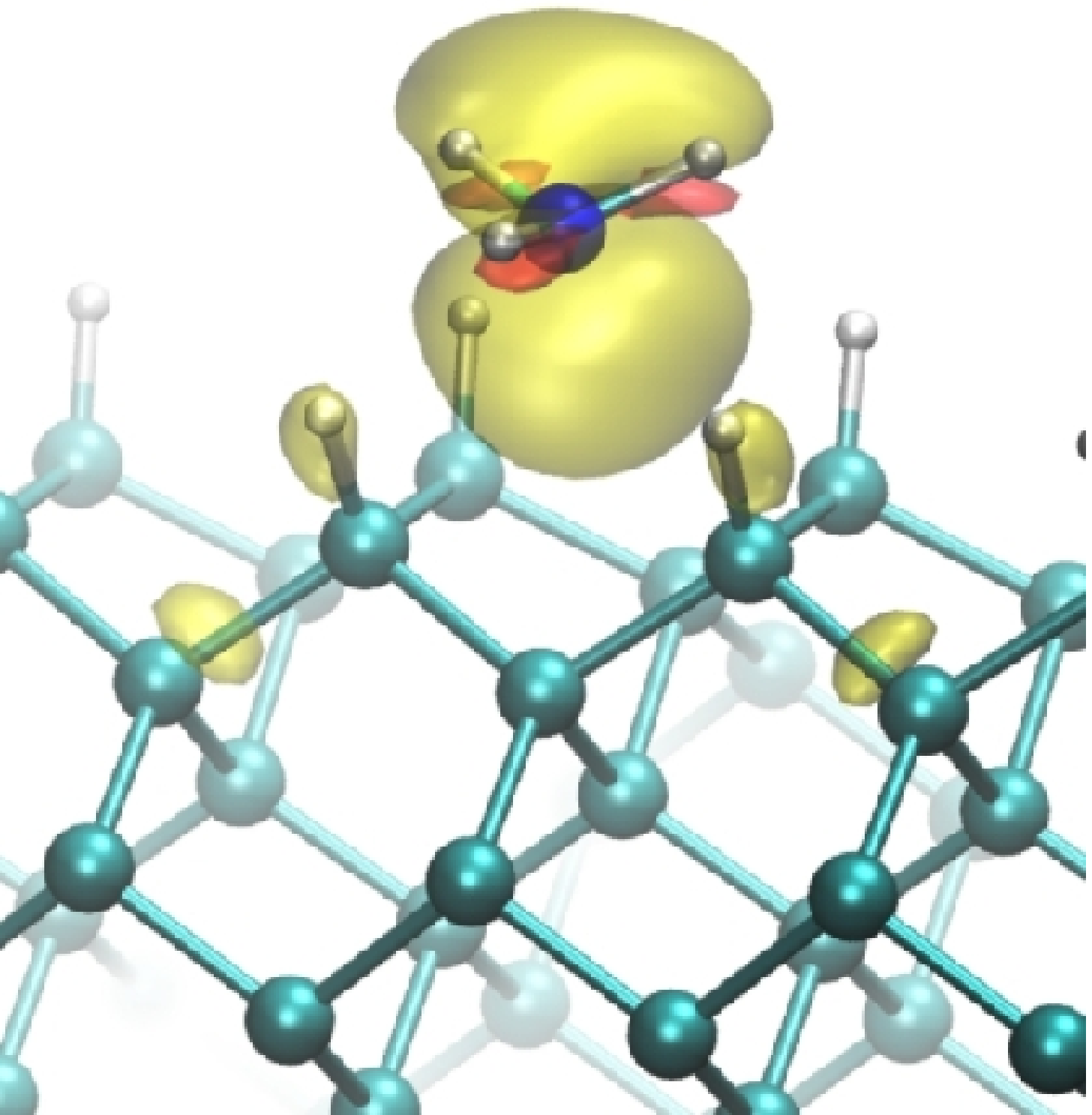}\tabularnewline
\end{tabular}
\end{figure*}

\begin{figure*}
\caption{\label{fig:casino}(color online) A synoptic scheme of the different migration
mechanisms we have considered. The energy of  local minima and 
transition states (in eV) are reported with reference to the desorbed SiH$_3$
(zero of energy). 
Activation energies for jumps in both directions can be obtained as energy differences. 
Transition-state energies for desorption from the physisorbed states,
either as silyl or as silane, are also reported (large arrows).
}
\centering\includegraphics[width=0.9\textwidth]{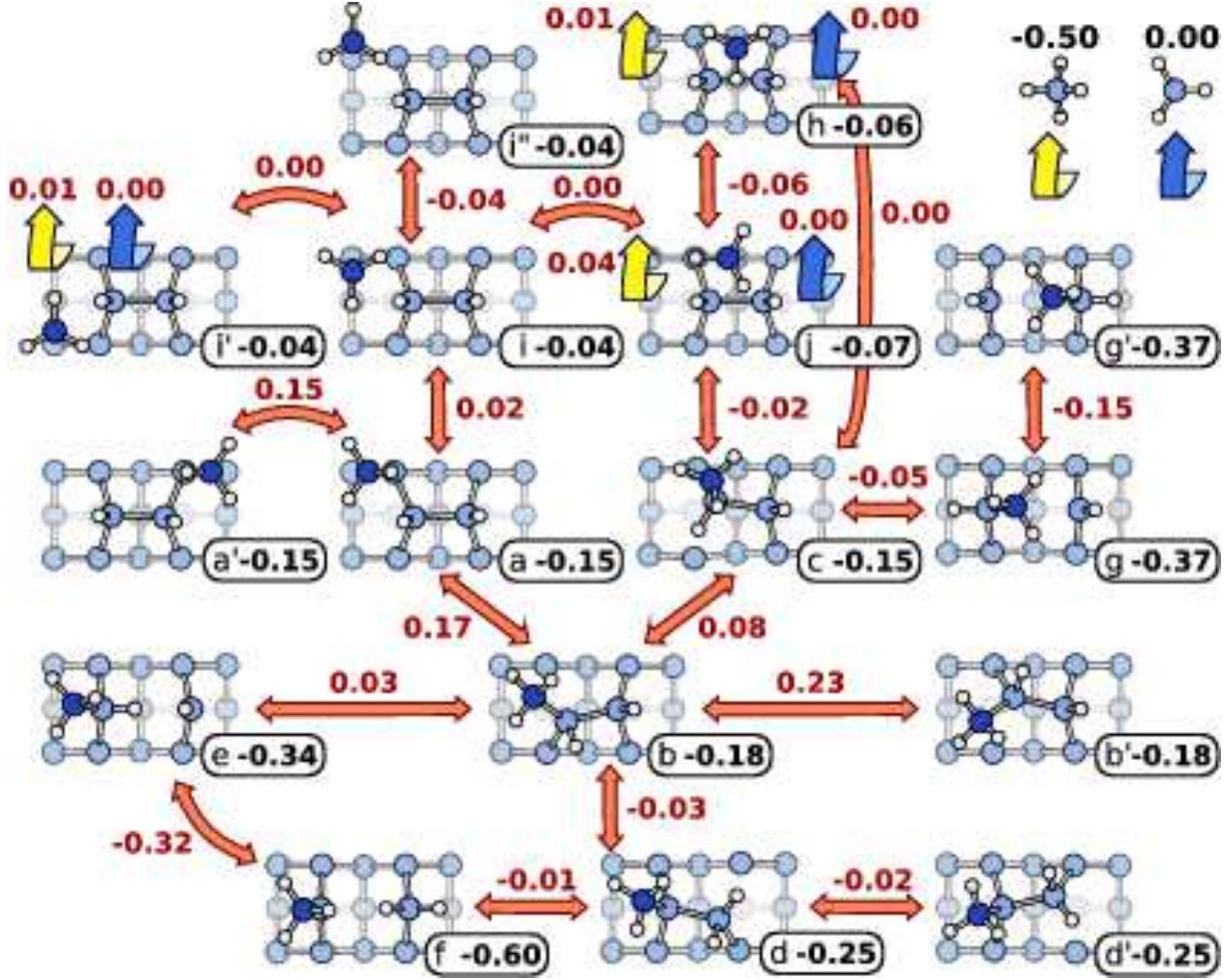}
\end{figure*}

Except for the overcoordinated configuration (a), the local minima  
 can be classified into three categories: configurations 
 with a broken dimer ((e), (f), (g)),
and the unpaired electron mostly localized on a surface $\mathrm{Si}$ atom,
configurations  with a broken backbond ((b), (c), (d)), and the unpaired electron
 localized onto a 
subsuperficial atom (see Figure~\ref{fig:spinloc}), and finally  weakly bound ``physisorbed'' 
configurations of the silyl radical ((h), (i), (j)).
Configurations in the first category appear to be lower in energy than  those in  the second one,
because the substrate is less strained and the $\mathrm{H}-\mathrm{H}$ repulsion is lower.

Configurations with adsorption energies in the range 0.15-0.35 eV are referred to hereafter as strongly bound
adsorption states, while traps are states with still higher adsorption energies. 
The three ``physisorbed'' states could be reached from  the gas phase with no activation barrier. 
However, because of the very tiny adsorption energy, adsorption at these sites might have a low cross
section. Indeed, these states have not been observed in our previous MD simulations of SiH$_3$ impinging on
the surface with energy in the range 01.-0.2 eV \cite{nostroprb}.
The existence of a precursor state for silyl 
adsorption on hydrogenated surfaces has been proposed in early models of PECVD growth\cite{matsuda1990},
although by assuming a physisorption geometry involving a three centers configuration ($\mathrm{H_3Si-H-Si_{surf}}$) which 
has been ruled out by 
later \emph{ab-initio} calculations on  $\mathrm{H:Si\left(111\right)}$ \cite{gupta2002} and
$\mathrm{H:Si\left(100\right)2\times 1}$ \cite{nostroprb} surfaces. 
The silyl can move from physisorbed states to  configurations with a higher
 adsorption energy by overcoming tiny barriers ($< 0.05$  eV) which can   
 be ascribed to hydride-hydride electrostatic repulsion. 


To complete the analysis of the adsorbed silyl, we have also considered 
desorption processes, either as $\mathrm{SiH_3}$ or as SiH$_4$, with the concurrent 
removal of a surface hydrogen. We have not been able to obtain a direct desorption mechanism 
from the  strongly bound adsorption states, as paths optimized with the NEB method always show the physisorbed states
as intermediates.
 Desorption as silyl from physisorbed configurations (h), (i) and (j) takes place with no
barrier other than the binding energy of the free radical. Energy barriers of
0.07, 0.05 and 0.11 eV have to be overcome by the silyl radical  to leave the surface as a silane molecule by removing a nearby hydrogen 
from configurations (h), (i) and (j), respectively.

To simplify the discussion of  Figure~\ref{fig:casino} which reports all the migration events
 we have investigated,
at first let us consider  jumps among strongly bound states.

\begin{figure*}
\caption{\label{fig:diffupaths}(color online) Diffusion pathways for a $\mathrm{SiH_3}$
radical, a) parallel to the dimer row, and b) perpendicular to the dimer row.
Energies of transition states and intermediate minima (in eV) are given. The zero of energy corresponds 
to the desorbed ${\rm SiH_3}$.}
\large\bfseries
\begin{tabular}[c]{@{}>{\raggedleft}b{0.05\textwidth}@{}>{\centering}m{0.95\textwidth}@{}}
a)\\\vfill &\includegraphics[width=0.9\textwidth]{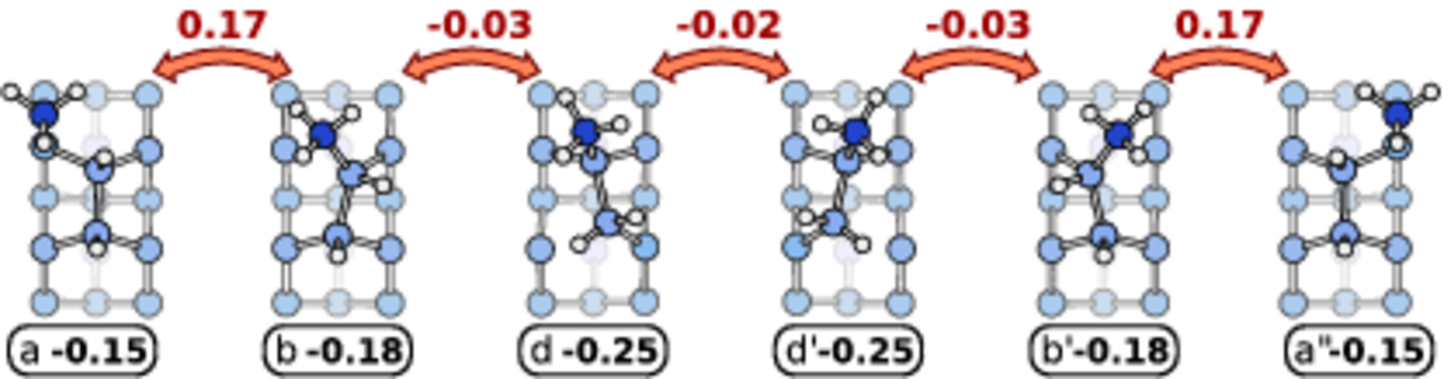}\tabularnewline
b)\\\vfill &\includegraphics[width=0.9\textwidth]{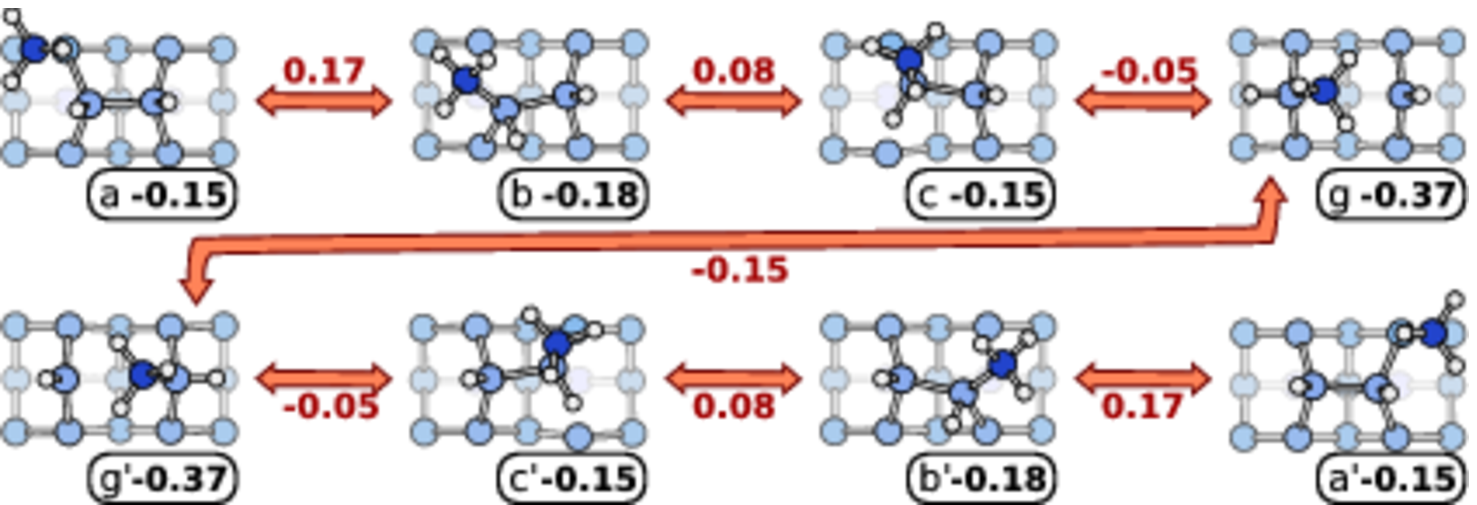}\tabularnewline
\end{tabular}
\end{figure*}

Starting from configuration (a),  we can identify  two diffusion pathways, 
one parallel to the dimer rows (Figure~\ref{fig:diffupaths} (a)) and  
the other in the perpendicular direction  (Figure~\ref{fig:diffupaths}b).
The overall activation barrier is 0.36 eV in both cases, corresponding to the 
key-event $\rm{(b)\rightarrow (a)}$. 
The direct 
(b)$\rightarrow$(b$'$) conversion has a higher barrier than a mechanism which 
takes place via reversible hydrogen exchange between the dimer atoms
((b)$\rightarrow$(d)$\rightarrow$(d$'$)$\rightarrow$(b$'$)), which was spotted by metadynamics.

\begin{figure}
\caption{\label{fig:traps}(color online) 
Two paths leading to the trap state (f) from the local minimum (b). The zero of energy (eV)  
corresponds to the desorbed SiH$_3$.}
\centering\includegraphics[width=0.9\columnwidth]{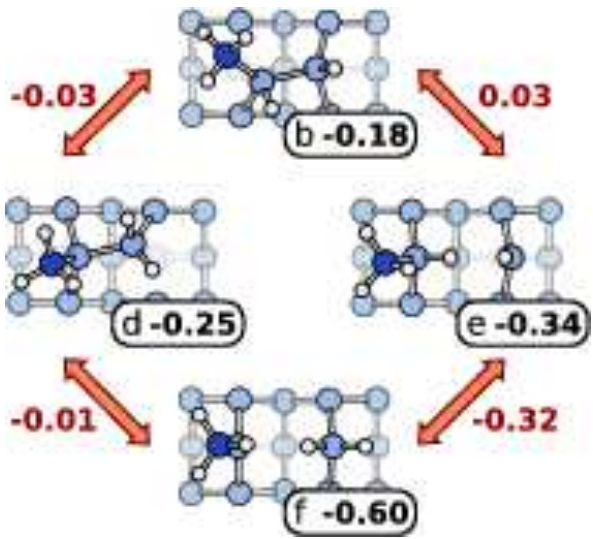}
\end{figure}

\begin{figure}
\caption{\label{fig:diffuphysi}(color online) 
A path for diffusion between two equivalent (a) sites, passing through physisorbed state (i).
Energies (eV) of minima and transition states are given. The zero of energy corresponds to the desorbed
 ${\rm SiH_3}$. See Figure~\ref{fig:minimi} for sketches of the
two structures viewed along $\left[0\bar{1}1\right]$.}
\centering\includegraphics[width=0.9\columnwidth]{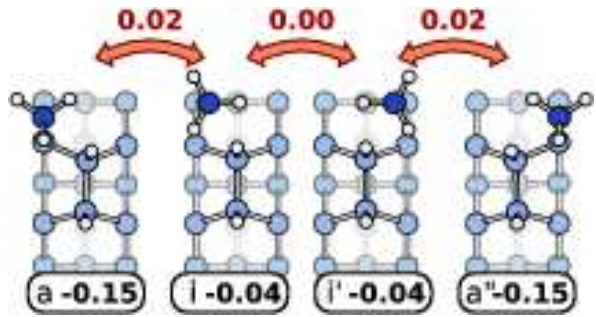}
\end{figure}

However, from configuration  (b), the radical can reach
the more stable state (f) state, following one of the two paths highlighted in Figure~\ref{fig:traps},
both of which have activation energies lower than the activation barrier for diffusion.
The deepest local minimum (f) acts as a ``trap'' state which hinders diffusion. 
The overall activation energy for diffusion from (f) site to an equivalent (f$'$) site along the dimer row
is as high as 0.77 eV.
Even though the presence of the trap state  rules out fast diffusion at low temperatures,
a barrier of 0.7 eV is still sufficiently low to be easily overcome  at room or growth 
temperatures. Still, we have not yet included the physisorbed states into
the picture. Actually,  some diffusion pathways involving  physisorbed 
states as intermediates have slightly lower barriers than the mechanisms involving 
 strongly bound states only, e.g. 
 the  $\rm{(a)\rightarrow (a')}$ path 
shown in Figure~\ref{fig:diffuphysi} 
with an overall barrier of 0.17 eV.

\begin{table}
\caption{\label{tab:attivaz} 
Activation energies (in eV) for various migration mechanisms of
$\mathrm{SiH_3}$ on $\mathrm{H:Si\left(100\right)-(2\times 1)}$. Column 
{\bf B} reports values obtained  by NEB optimization of the  minimum energy path
by  sampling the BZ at the $\Gamma$-point only. 
Column {\bf A} corresponds to the activation energies computed with one special k-point on
geometries optimized with $\Gamma$-point only.
The primed and not-primed configurations are  equivalent by symmetry: 
$\rm{(a)\rightarrow (a')}$ corresponds to a jump across the channel,
$\rm{(b)\rightarrow (b')}$ and $\rm{(d)\rightarrow (d')}$ corresponds to the conversion 
between two structures equivalent with respect to the mirror plane 
orthogonal to $\left[0\bar{1}1\right]$,
$\rm{(g)\rightarrow (g')}$ corresponds to the migration of the radical between
the two atoms of the broken dimer, 
$\rm{(i)\rightarrow (i')}$ and $\rm{(i)\rightarrow (i'')}$ correspond to 
two steps of a jump along the dimer row, between two
equivalent physisorbed states.}
\begin{ruledtabular}
\begin{tabular}[c]{ l c c | l c c }
Mechanism	& {\bf A} 	&{\bf B} 	& Mechanism 	& {\bf A} 	&{\bf B} 	\\
\hline
$\rm{a\rightarrow a'}$&	0.30	& 0.35		&$\rm{c\rightarrow j}$&	0.14	& 0.27	  	\\
$\rm{a\rightarrow b}$&	0.32	& 0.37		&$\rm{d\rightarrow d'}$&0.22	& 0.27	  	\\
$\rm{a\rightarrow i}$&	0.17	& 0.24		&$\rm{d\rightarrow f}$&	0.24	& 0.39	  	\\
$\rm{b\rightarrow b'}$&	0.42	& 0.55		&$\rm{e\rightarrow f}$&	0.02	& 0.03	  	\\
$\rm{b\rightarrow c}$&	0.26	& 0.37		&$\rm{g\rightarrow g'}$&0.22	& 0.35	  	\\
$\rm{b\rightarrow d}$&	0.15	& 0.28		&$\rm{h\rightarrow j}$&	$<$0.01	&$<$0.01	\\
$\rm{b\rightarrow e}$&	0.22	& 0.37	  	&$\rm{i\rightarrow i'}$&0.04	& 0.03	  	\\
$\rm{c\rightarrow g}$&	0.10	& 0.21		&$\rm{i\rightarrow i''}$&$<$0.01	&$<$0.01 	\\
$\rm{c\rightarrow h}$&	0.15	& 0.27		&$\rm{i\rightarrow j}$&	0.05	& 0.04	 	\\
\end{tabular}
\end{ruledtabular}
\end{table}

\begin{table}
\caption{\label{tab:modelkmc} Energy (eV) of  ``minima'' and ``transition states''
used in the simplified Kinetic Monte Carlo simulation based on  Figure~\ref{fig:simple}.
The zero of energy corresponds to the desorb silyl (``silyl'' in the table).
For simplicity we label as ``chemisorbed'' (here and in Fig. \protect\ref{fig:simple})
the bound states of SiH$_3$ with adsorption energy in the range 0.15-0.35 eV.
Prefactors for the ``fixed-prefactor'' model have been set equal to 1~THz, or twice as large in the 
presence of two symmetry-equivalent paths.
We have also considered rate prefactors $\nu^\star$ obtained with harmonic transition state theory
for selected processes, namely (a)$\rightarrow$(b) for the reactions starting in a 
strongly bound state (``chemi'' or ``trap'' in the table), (i)$\rightarrow$(i')
for reactions starting in a physisorbed state, and (i)$\rightarrow \rm SiH_3$
for desorption. The ``forward'' and ``backward'' values for $\nu^*$ (THz) 
 used in the simulations of Figure~\ref{fig:kmcprefix} are given.
}

\begin{ruledtabular}
\begin{tabular}[c]{ l c | l c c c}
State	& Energy& State	& Energy & $\nu^{\star}_{\rightarrow}$ & $\nu^{\star}_{\leftarrow}$\\
\hline
trap	& -0.59 & trap-chemi	& 0.09	& 56	& 28\\
chemi	& -0.15	& chemi-physi	& -0.01	& 28	& 2\\
physi	& -0.05 & chemi-chemi	& 0.15	& 28	& -\\
silyl	& 0.00  & physi-physi	& 0.00	& 2	& -\\
silane	& -0.50 & physi-silyl	& 0.00	& 0.8	& -\\
	&	& physi-silane	& 0.00	& 0.8	& -\\
\end{tabular}
\end{ruledtabular}
\end{table}

\begin{figure}
\caption{\label{fig:simple}(color online) 
Simplified scheme that accounts for most of the features of the complex reactions scheme
of Figure~\ref{fig:casino}; the activation energies have to be 
considered just qualitative estimates,  because of both the limited accuracy of DFT
calculations and the fact that each of the barriers in this figure corresponds to different 
pathways with slightly different activation energies (cf. Figure~\ref{fig:casino}).
For simplicity we label as ''chemisorbed'' (here and in Table \protect\ref{tab:modelkmc})
the bound states of SiH$_3$ with adsorption energy in the range 0.15-0.35 eV.}
\centering\includegraphics[width=0.9\columnwidth]{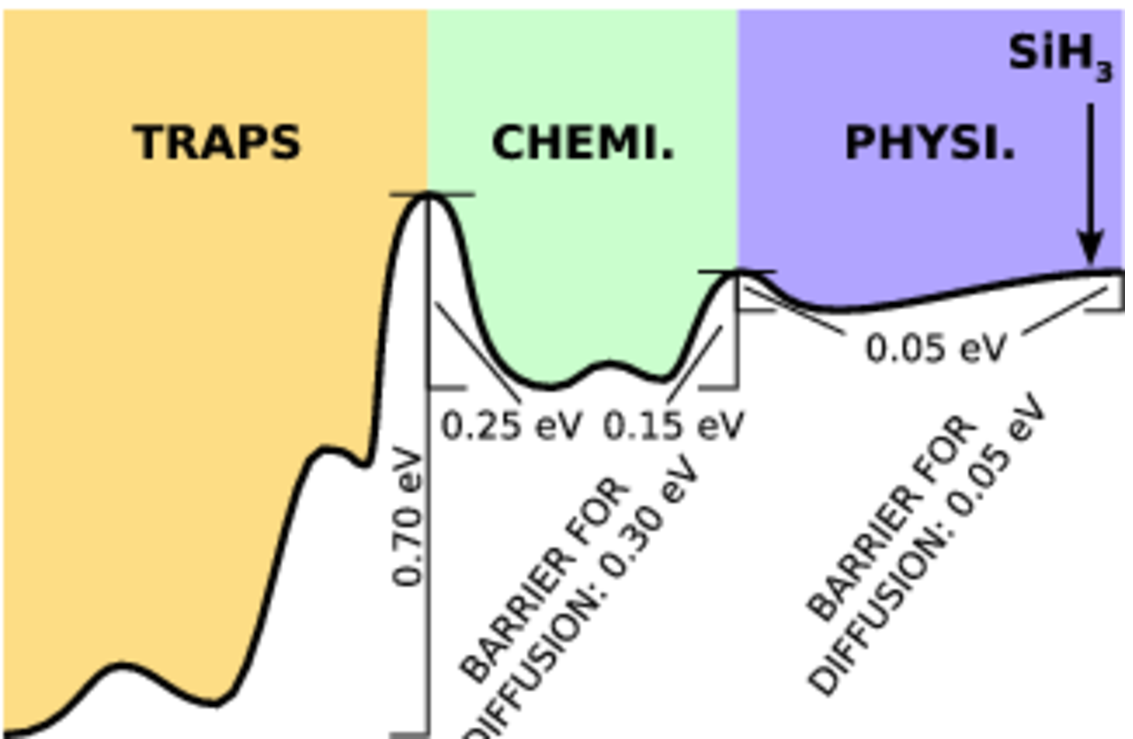}
\end{figure}

The evolution of the adsorbed SiH$_3$ radical we have discussed so far
 can be summarized by  the simplified
scheme of Figure~\ref{fig:simple}. A high barrier has to be overcome to escape from the realm 
of trap states 
and reach some ``mobile'' local minima. From one of those states, 
with similar barriers,
the system might fall back into the traps, diffuse to nearby strongly bound states or reach 
one of the physisorbed states.
Barriers for migration between
different strongly bound states ($\sim$ 0.3 eV) are similar to barriers for migration between strongly bound and 
physisorbed states.
From the latter physisorbed states, with barriers below 0.1~eV,
the silyl can go back to the strongly bound states, diffuse amongst the physisorbed states
or desorb, with similar barriers, either as silyl or as silane. 
The accuracy of our calculations is not sufficient to identify which are the dominant processes 
at low temperature,  since uncertainties of the order of 0.1 eV in the activation energies  are expected due
to approximations in   the exchange and correlation functional. However,
at typical growth temperatures of PECVD and LEPECVD the different events are likely to have
similar probability.
To estimate the average diffusion length that a silyl radical can travel before desorption takes place,
we have performed kinetic Monte Carlo simulations using the activation energies for the different processes
summarized in Table III. The results are discussed in the next section.

As a final remark, we  compare our results on diffusion mechanisms with previous ab-initio works appeared in
literature.
The migration 
$\rm{(a)\rightarrow(a')}$ between two adjacent dimers rows and the $\rm{(g)\rightarrow(g')}$ 
jump within an open dimer have been investigated  in Ref.\cite{bakos2006} within a framework very similar
to ours which, however, produced activation energies as high as
0.9 and 0.6 eV for the two process, respectively, as opposed to our values of 0.3 and 0.22 eV. 
The discrepancy is due  to a very different geometry of the transition state which 
in Ref.\cite{bakos2006} corresponds (both for $\rm{(a)\rightarrow(a')}$ and $\rm{(g)\rightarrow(g')}$ jumps)
to a silyl nearly flat, $sp^2$-like followed by
 an ``umbrella flip'' of the hydrogen atoms of the silyl. In our case, 
by choosing 
the trajectory obtained from the metadynamics simulation 
as initial MEP in the NEB optimization, 
we have been able to find a lower transition state involving  small deformation
of the $\mathrm{SiH_3}$ radical
which undergoes a  rotation around its $C_3$ axis during the jump still keeping a  $sp^3$-like conformation.
Indeed, within the NEB method, it is possible that an unappropriate choice of the initial MEP 
might lead to a reaction path with an activation energy higher than the lowest energy
path to the products.
Actually,  we remark that an even lower barrier (0.17 vs 0.30 eV) for the jump across the channel between two
neighboring rows 
$\rm{(a)\rightarrow (a')}$ can be obtained along the path  
$\rm{(a)\rightarrow (i)\rightarrow (a')}$, 
with the physisorbed state (i) as an intermediate.
Another issue worth being mentioned is a discrepancy we have found
with previous work  \cite{bakos2006} on the diffusion mechanism of the silyl 
along the dimer row.
Using LSD-PBE (spin unrestricted calculations) 
 we have not been able to optimize the intermediate state described as a local minimum
in Figure~3 of Ref.\cite{bakos2006} (a silyl adsorbed on the five-fold coordinated Si atom of the Si-H group)
which has been  obtained probably within a spin-restricted (LDA-PBE) framework.
In fact, the configuration proposed as a local minimum  along the diffusion path
in Ref. \cite{bakos2006},
transforms in  configuration (e), in our NEB optimization. 
Moreover, by searching for                                
 a direct (a)$\rightarrow$(e) transformation, we have always found configuration (b) as an intermediate minimum.
Consequently,   we have not been able
to find a minimum energy path for a ``direct'' jump $\rm{(a)\rightarrow (a'')}$ along
the dimer row as reported in Ref. \cite{bakos2006}.
Probably
the discrepancies with previous work might still be  ascribed to the neglect of spin polarization 
in Ref. \cite{bakos2006}.

\subsection{Kinetic Monte Carlo}

Based on  the reaction scheme of Figure~\ref{fig:casino} we have performed Kinetic Monte
Carlo (KMC) simulations of diffusion and desorption of the adsorbed silyl.
 We have considered a lattice model, each lattice site corresponding to a
$2\times 1$ unit cell. At each site different configurations for the silyl, 
corresponding to the different minima of Table II (and Fig. ~\ref{fig:casino}) are possible. 
We have only omitted configuration (e), 
as it is only a shallow minimum along the  $\rm{(b)\rightarrow (f)}$ migration pathway,
causing  frequent oscillation of the system between minima (f) and (e)  which slows down 
dramatically simulations at low (room) temperatures.
We have estimated reaction rates $\nu$
at different temperatures within transition state theory, i.e.  
$\nu=\nu^\star \exp\left(-E_a/k_BT\right)$, with activation energies $E_a$ given in
Table III and  prefactor $\nu^\star$  set to 
 1~THz for all processes. In some cases, 
for example for the $\rm{(a)\rightarrow (a')}$ jump, two pathways equivalent by 
symmetry exists: we have used a 2~THz prefactor in this and similar cases.
 The choice of a common prefactor is obviously questionable. 
We have calculated within the harmonic approximation to transition state theory the 
prefactors for a diffusion process among strongly bound 
states ((a)$\rightarrow$(b)), for the diffusion between two 
physisorbed states ((i)$\rightarrow$($\rm i'$)) and 
for desorption ((i)$\rightarrow$$\rm SiH_3$).
The prefactor $\nu^\star$ is given by $\prod \nu^{(I)}_j /\prod \nu^{(TS)}_j$, 
where $\nu^{(I)}_j$ and $\nu^{(TS)}_j$ are the positive phonon frequencies 
for the initial and transition states respectively,
obtained by diagonalizing the dynamical matrix, 
calculated in turns by finite displacements of the atoms. 
For the desorption process, which does not display a saddle point on the potential energy surface
   along the desorption path,
we have instead used the vibrational frequency of the phonon whose displacement pattern
is parallel to the minimum energy path.
The results, $28$~THz ((a)$\rightarrow$(b)), $2$~THz ((i)$\rightarrow$($\rm i'$)) and 
$0.8$~THz ((i)$\rightarrow$$\rm SiH_3$), differ less than the typical 
errors in this kind of calculations, which are at least one or two orders of magnitude.
Therefore, given the considerable computational resources needed to compute all prefactors,
we have judged not worthy to pursue this task at the moment.

Moreover, we have considered
 a single silyl adsorbed on a perfect H:Si(100)-(2x1) surface which is
obviously very far from the real growth conditions in PECVD where partial surface hydrogenation and 
large coverage of adsorbed species are expected. 
Still, these simplified KMC simulations can provide crucial information on SiH$_3$ diffusion
to model the behavior of the silyl radical at the more complex conditions of real PECVD growth. 

Within our scheme, the final fate of the adsorbed silyl is always desorption, either as 
a silyl or a silane. 
The KMC simulations let us estimate the average resident time of the silyl on the surface and the
average maximum distance the silyl can travel before desorbing.
We have checked the stability of the results versus random variations
of the barriers within the typical DFT uncertainties. To this aim, we have 
performed several sets of simulations adding a uniform noise with zero mean and $\pm 0.1$~eV
width to all the minima and transition states energies in Figure~\ref{fig:casino}, discarding 
the set of randomized activation energies whenever  one of them is negative.
Some other sets have also been discarded  since they  lead the  system to oscillate for
 a long time ($>10^6$ steps) between two minima separated by tiny barriers. 
In all simulations the silyl starts from site (a) 
since it corresponds to the adsorption site  observed in dynamical Car-Parrinello
simulations of SiH$_3$ impinging on the surface (cf. Section IIIA).

\begin{figure}
\caption{\label{fig:escapetime}
Distribution of  resident times $t$ of the silyl on the surface (before desorption).
The initial configuration is always site  (a) (see text). 
The results are averaged over $10^6$ independent KMC runs.
The distribution is shown with a logarithmic scale on abscissa ($h(x), x=log_{10} t$).
Panel a) and b) correspond to simulations at $T=500$~K and
 $T=1000$~K, respectively. Curve {\bf A} corresponds to the
full reaction scheme, with the energies given in Figure~\ref{fig:casino}.
Curve {\bf B} corresponds to the complete scheme, where for each simulation
energies for the minima and the transition states have been randomized (see text).
 Curve {\bf C} and {\bf D} corresponds to the simplified model,
with  parameters  given in Table~\ref{tab:modelkmc} and or with 
randomized energies (see text), respectively.}
\large\bfseries
\begin{tabular}[c]{@{}>{\raggedleft}b{0.05\columnwidth}@{}>{\centering}m{0.95\columnwidth}@{}}
a)\\~\\~\\  & \includegraphics[width=0.9\columnwidth]{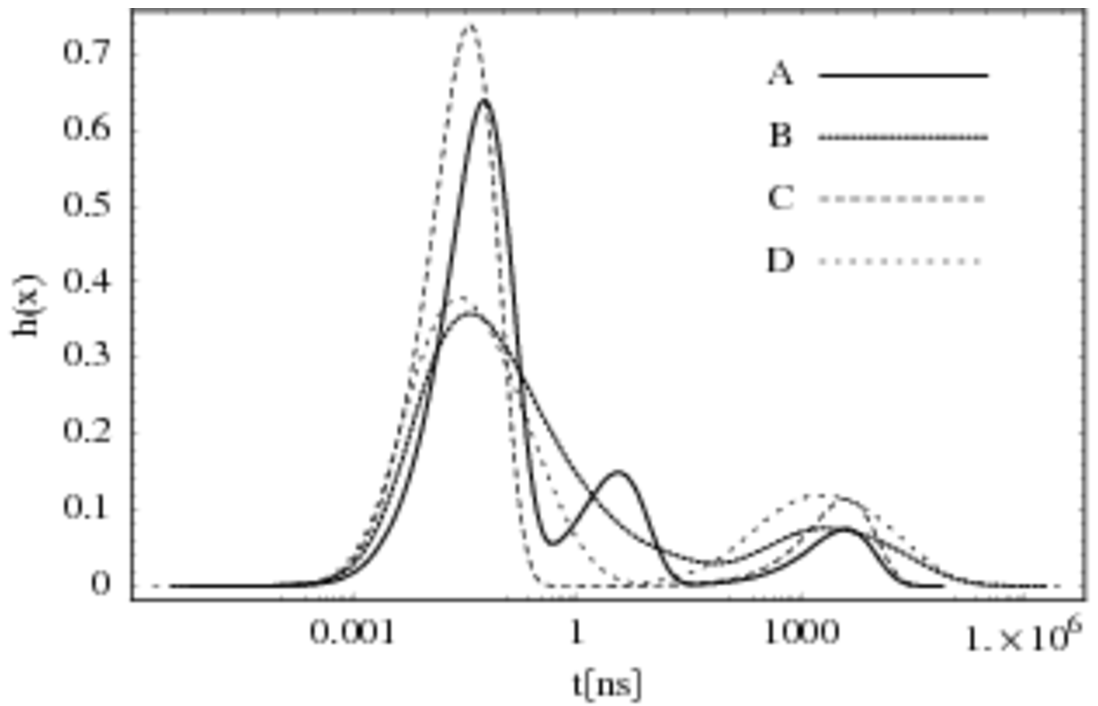}\tabularnewline
b)\\~\\~\\  & \includegraphics[width=0.9\columnwidth]{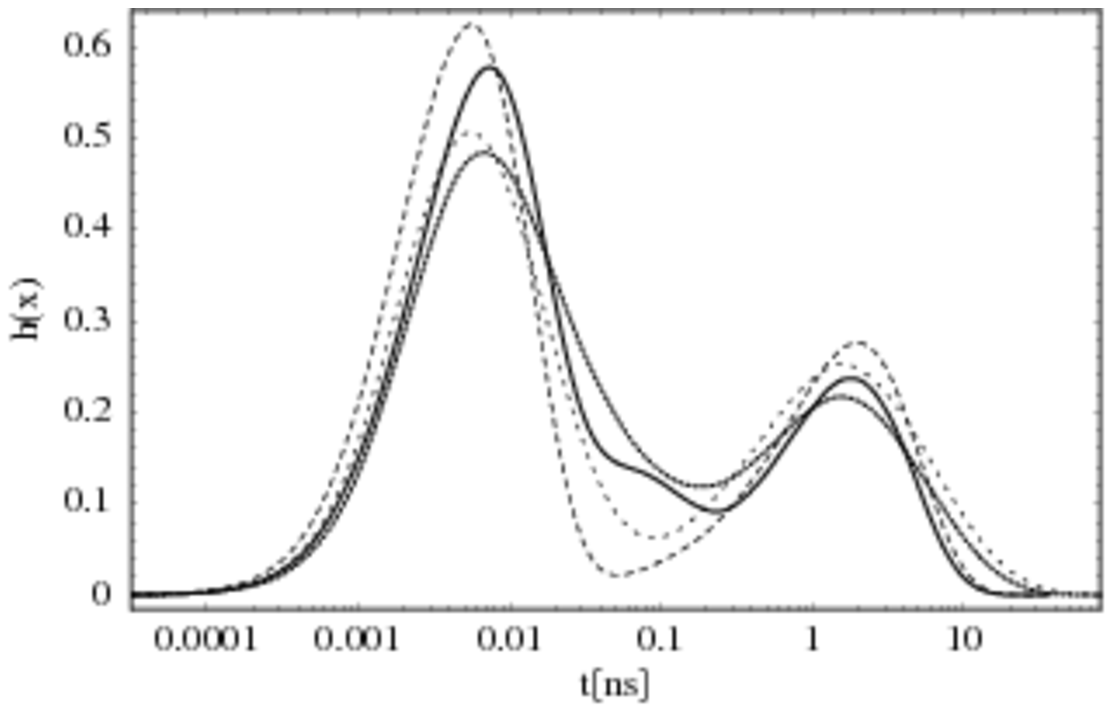}\tabularnewline
\end{tabular}
\end{figure}

The calculated distribution of resident times  (times before desorption) 
is reported in Fig. \ref{fig:escapetime}, as obtained by averaging over $10^6$ KMC simulations.
Results with activation energies from Table III and with the energies randomized in every simulation  are compared.
The distribution of resident times, $h(x)$, is reported as a function of the logarithm of the
resident time $t$ ($x=log_{10}t$), i.e. $\int_{x_1}^{x_2} 
h\left(x\right) {\rm d}x$ returns the fraction of trajectories with a resident time
between $10^{x_1}$ and $10^{x_2}$. In this representation,  for an exponential decay $ke^{-kt}$,
the  histogram on a logarithmic $x$-axis has a peak at 
$-\log_{10} k$. Different peaks on the histogram are therefore related to processes with 
different desorption rates.
The distribution of resident times  reveals three peaks at 500 K: a nearly instantaneous
desorption from site (a) (low $t$ peak), a desorption from site (g) easily accessible from (a)
(peak at intermediate $t$) and finally desorption from trap states at longer $t$.
The peak due to desorption from site (g) reduces to a shoulder of the first peak in  simulations
with randomized activation energies.
At higher temperatures the probability to reach the traps increases and the importance of longer
resident times (area of the peak at the longest $t$) increases.
Therefore, most of the radicals desorb quickly, without reaching the trap states, but a fraction 
of the adsorbed ${\rm SiH_3}$ get to the trap, and remain adsorbed for longer times,
of the order of $10^{-5}$~s at $500$~K.
The absolute position of the peaks in $h(x)$ depends on the choice of the prefactor
$\nu^*$ in the reaction rates, but the ratio of different positions and the relative area 
of the peaks do not. 

\begin{figure}
\caption{\label{fig:distmax}
Maximum distance along $\left[0\bar{1}1\right]$ reached by the silyl before desorption
as a function of temperature. 
The results are averaged over $10^6$ independent KMC runs. 
Curve {\bf A} corresponds to the
full reaction scheme, with the energies given in Figure~\ref{fig:casino}.
Curve {\bf B} corresponds to the complete scheme with randomized activation energies (see text).
 Curve {\bf C} and {\bf D} corresponds to the simplified model,
with  parameters  given in Table~\ref{tab:modelkmc} and or with            
randomized energies (see text), respectively.
Distances are given in units of the lattice spacing along $\left[0\bar{1}1\right]$
($a_y$=3.84 \AA).
}
\includegraphics[width=0.9\columnwidth]{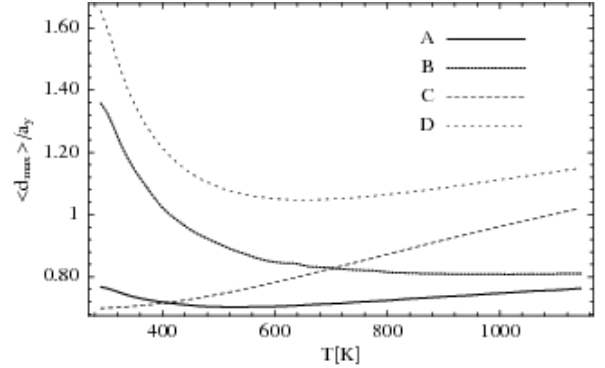}\tabularnewline
\end{figure}

\begin{figure}
\caption{\label{fig:kmcprefix}
Comparison of KMC simulations performed on the simplified model and constant
prefactors (continuous curve {\bf A}) and with prefactors chosen according to 
harmonic transition state theory predictions (dotted curve {\bf B}, cf. 
Table~\ref{tab:modelkmc}). In both cases averages are performed over 
$10^6$ independent KMC runs, with activation barriers randomized as 
described in the text. 
a) Distribution of  resident times of 
the silyl on the surface at 500~K 
(cf. Figure~\ref{fig:escapetime}).  b) Maximum distance along 
$\left[0\bar{1}1\right]$ reached by the silyl before desorption as a 
function of temperature (cf. Figure~\ref{fig:distmax}).
Distances are given in units of the lattice spacing along $\left[0\bar{1}1\right]$
($a_y$=3.84 \AA).
}
\large\bfseries
\begin{tabular}[c]{@{}>{\raggedleft}b{0.05\columnwidth}@{}>{\centering}m{0.95\columnwidth}@{}}
a)\\~\\~\\ & \includegraphics[width=0.9\columnwidth]{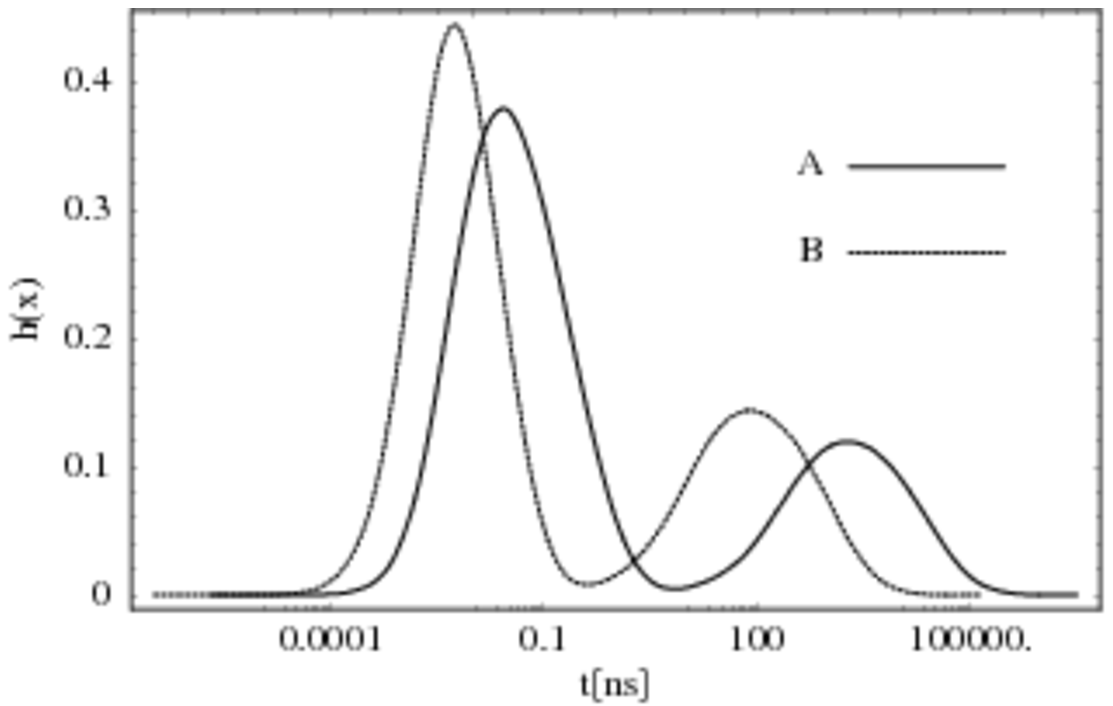}\tabularnewline
b)\\~\\~\\  & \includegraphics[width=0.9\columnwidth]{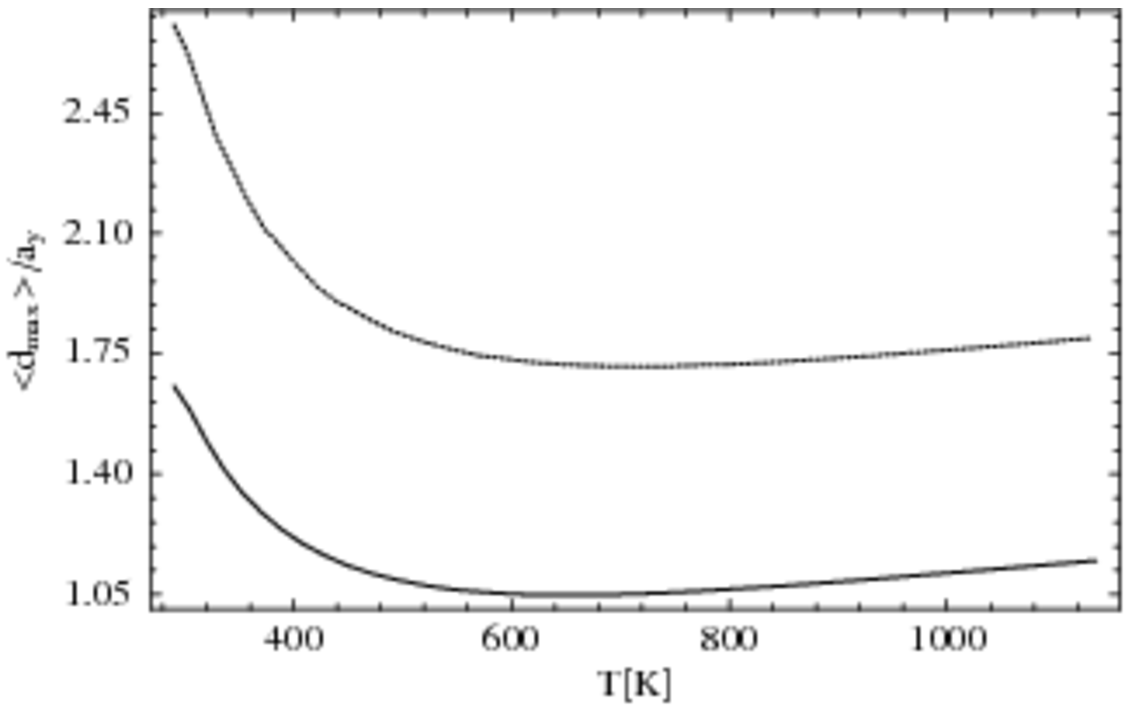}\tabularnewline
\end{tabular}
\end{figure}

In Figure~\ref{fig:distmax} we report the results on the average maximum distance
 the radical can travel before desorbing. For simplicity, only  displacement
along $\left[0\bar{1}1\right]$ are considered.
 Diffusion distances
have been calculated as the displacement along $\left[0\bar{1}1\right]$ from the initial
site (a); each  minimum within the unit cell
corresponds to a displacement equal to the relative coordinate  
of the ${\rm Si}$ atom with respect to site (a). 
Averaging over random changes of the barriers results in 
a variation of the relative importance of diffusion and desorption processes
which is significative at room temperature, but negligible at higher temperatures. This is easy to 
understand, as, at room temperature, the radical would travel very long distances if the barrier
for desorption was even a few tens of meV higher than that for diffusion among 
physisorbed states. Anyway, it is very unlikely that the radical could jump further than
 few cells before desorbing. 
The average diffusion lengths in Fig. \ref{fig:distmax} are independent from the value chosen
for the common rate prefactor $\nu^*$.

To investigate the dependence of the diffusion length on the variation of rate prefactors
 we have considered a simplified model corresponding
to  Figure~\ref{fig:simple}, with the set of parameters reported in Table~\ref{tab:modelkmc}. Although the 
model is much simpler than the complete reaction scheme, there are still too many parameters 
to attempt a significative fitting procedure. We have thus chosen the values in 
Table~\ref{tab:modelkmc}
heuristically, as we only wish to show that the complete scheme is very redundant, and the
simplified scheme provides a reasonable description of the behavior of the silyl.
In the simplified model only jumps between adjacent cells are considered. Displacement of the silyl
among different minima within the same unit cell are not included in the calculation of the
maximum average displacement. 
Firstly, we set all the prefactors equal, as done in the fully detailed KMC, for sake of
comparison. The results are reported
in Figs. \ref{fig:escapetime} and \ref{fig:distmax}. 
The simplified model is able to reproduce the main features of the complete KMC results. 
 Both the shape of $h(x)$ (but for the smaller structure due to desorption from (g))
and the order of magnitude of the average diffusion length are well reproduced. The simplified model
can thus be considered as a good starting point to develop a more elaborate KMC model for
realistic conditions with variable H and SiH$_3$ coverage.  
Then, we have investigated  the dependence of the results on the choice of different
prefactors for different reactions. To this aim we have used the prefactors obtained
by harmonic transition state theory for the three representative reactions (Table~\ref{tab:modelkmc}). 
The residence-time histogram (Figure~\ref{fig:kmcprefix}) is only slightly shifted. 
The attempt frequency for diffusion is in this case almost three times the one 
for desorption, but the average diffusion length is only doubled.
This suggests that the model is also relatively stable against changes in the 
estimated prefactors, making our predictions reliable, despite the unavoidable errors
on energies and overall reaction rates.
We can therefore conclude that desorption events prevents diffusion of $\rm SiH_3$ radicals 
for more than a few lattice spacings. Silyl migration can account for local rearrangement
effects, resulting in an increase in the capture area of surface dangling bonds, but cannot
be invoked to explain long-range diffusion or surface smoothening on a mesoscopic scale.

\section{Conclusions}

We have investigated the diffusion and desorption mechanisms of the SiH$_3$ radical adsorbed
on H:Si(100)-(2x1) by means of ab-initio calculations.
Preliminary metadynamics Car-Parrinello simulations aided us in identifying local minima
and diffusion pathways of the the adsorbed silyl.
Activation energies and minimum energy paths for diffusion and desorption have then been
refined by NEB calculations.
We have identified three classes of adsorption geometries: a set of physisorbed states with low
adsorption energy ( $\sim$ 5 meV), more  strongly bound states with adsorption energies in the range
0.15-0.35 eV and a trap state with adsorption energy of 0.6 eV.
The silyl can escape from the trap state and diffuse among the other more strongly bound minima.
However, by overcoming similar barriers the silyl can move to the physisorbed states from which it can
easily desorb.
Kinetic Monte Carlo simulations based on the full ab-initio reactions scheme show that the silyl radical
can diffuse on average by  a few lattice spacing before desorbing  in the temperature range
300-1000 K which includes  LEPECVD conditions.
In contrast with previous works \cite{valipa2005prl}, we conclude that diffusion of SiH$_3$ 
on
H:Si(100)(2x1) surface over length
longer than a few lattice spacing  is preempted by desorption.  
Therefore, fast diffusion  from H-rich to H-poor regions  
is not supposed to play a role in promoting the decomposition of SiH$_3$. 
Instead, at the experimental conditions of LEPECVD, 
 etching of some surface hydrogen atoms  by impinging SiH$_3$ or H atoms in the plasma
is presumably the key factor which promotes decomposition and insertion of SiH$_3$ adsorbed on a H-rich
surface region \cite{nostroprb}.

\section{Acknowledgments}

Discussion with  C. Cavallotti, S. Cereda, L. Miglio and F. Montalenti are gratefully acknowledged.
This work is partially supported by the Cariplo Foundation (SIMBAD project).
Computational resources at CINECA has been provided by CNISM
through "Iniziativa Calcolo Parallelo 2006".

\bibliographystyle{apsrev}

\cleardoublepage


\end{document}